\documentclass[aps,pra,twocolumn,notitlepage,superscriptaddress,10pt,floatfix]{revtex4-1}
\usepackage[pdftex,plainpages=false,colorlinks=true,citecolor=blue,linkcolor=blue,urlcolor=blue,filecolor=green,bookmarksopen=true]{hyperref}
\usepackage[utf8]{inputenc}
\usepackage{amsmath}
\usepackage[caption = false]{subfig}
\usepackage{graphicx,epstopdf}
\usepackage{blindtext}
\usepackage{lipsum}
\usepackage{amsfonts}
\usepackage{bbm}
\usepackage{amssymb}
\usepackage{enumerate}
\usepackage{color}
\usepackage{latexsym}
\usepackage{times,txfonts}
\newtheorem{theorem}{Theorem}

\newtheorem{definition}[theorem]{Definition}

\newcommand{\Cmath}{\mathcal{C}}

\newcommand{\Hcal}{\mathcal{H}}
\newcommand{\Gcal}{\mathcal{G}}
\newcommand{\Dcal}{\mathcal{D}}

\newcommand{\Lcal}{\mathcal{L}}
\newcommand{\Ecal}{\mathcal{E}}
\newcommand{\Rcal}{\mathcal{R}}

\newcommand{\Ocal}{\mathcal{O}}
\newcommand{\Fcal}{\mathcal{F}}

\newcommand{\Vcal}{\mathcal{V}}

\newcommand{\1}{\mathbbm{1}}
\newcommand{\Lmath}{\mathbbm{L}}

\newcommand{\dket}[1]{| #1 \rangle\!\rangle}
\newcommand{\dbra}[1]{\langle\!\langle #1 |}
\newcommand{\ket}[1]{| #1 \rangle}

\newcommand{\dinterpro}[2]{\langle\!\langle #1 | #2 \rangle\!\rangle}
\newcommand{\interpro}[2]{\langle #1 | #2 \rangle}
\newcommand{\bra}[1]{\langle #1 |}
\newcommand{\tr}[1]{ \text{Tr}\left\{ #1 \right\}}
\newcommand{\trs}[1]{ \text{Tr} \{ #1 \}}

\usepackage[dvipsnames]{xcolor}
\definecolor{CommentCollor}{HTML}{f25c54}
\definecolor{CancelCollor}{HTML}{a21112}
\definecolor{MyCollor}{HTML}{005f73}

\newcommand{\UFSCar}{Departamento de Física, Universidade Federal de São Carlos,\\ Rodovia Washington Luís, km 235 - SP-310, 13565-905 São Carlos, SP, Brazil}

\begin{document}

\title{Generalized transitionless quantum driving for open quantum systems}

\author{Alan C. Santos}
\email{ac\_santos@df.ufscar.br}
\affiliation{\UFSCar}

\author{Marcelo S. Sarandy}
\email{msarandy@id.uff.br}
\affiliation{Instituto de F\'{i}sica, Universidade Federal Fluminense, Av. Gal. Milton Tavares de Souza s/n, Gragoat\'{a}, 24210-346 Niter\'{o}i, Rio de Janeiro, Brazil}

\begin{abstract}
A general approach for transitionless quantum driving in open quantum systems is introduced. Under the assumption of adiabatic evolution for time-local master equations, 
we derive the generalized transitionless Lindbladian required to implement a shortcut to adiabaticity in an open system scenario. The general counter-diabatic Lindbladian obtained accounts 
for a phase freedom, which translates into a set of free parameters throughout the dynamics. We then discuss how our generalized approach allows us to recover 
the transitionless Lindbladian introduced by G. Vacanti~\textit{et al.} [New J. Phys. \textbf{16}, 053017 (2014)].  We then show how to engineer time-independent 
master equations that provide the same dynamics as the time-dependent master equation provided by the standard transitionless quantum driving in open systems. 
We illustrate our results by applying them both to the adiabatic Deutsch algorithm under dephasing and to the Landau-Zener Hamiltonian under bit-phase-flip. 
\end{abstract}

\maketitle

\section{Introduction}

Inverse quantum engineering is a useful approach to drive quantum systems through some desired path in parameter space and, hence, achieve a target 
final state~\cite{Chen:11,Jing:13,Kang:16,Yu:18,Santos:18-a,Chen:18-1,Wu:21}. Within a number of different approaches for inverse engineering, one can highlight the 
adiabatic dynamics as an important strategy, with successful applications in quantum control~\cite{Kral:07,Odelin:19} and quantum computation~\cite{Farhi:01,Tameem:18}. 
However, the requirement of a sufficiently large evolution time may lead the system to undesired phenomena due to decoherence~\cite{Childs:01,Sarandy:05-1,Jing:16,Albash:15,Venuti:17}. 
This has strongly motivated the investigation of methods for speeding up the adiabatic process (more precisely, to mimic the adiabatic behavior). 
In this scenario, transitionless quantum driving (TQD)~\cite{Demirplak:03,Demirplak:05,Berry:09} has been established as a widely used method for yielding 
shortcuts to adiabaticity, where additional fields are used to inhibit any diabatic transition between energy levels of the Hamiltonian. 
TQD has provided numerous applications in different branches of  
physics~\cite{Ruschhaupt:12,Adolfo:14,Stefanatos:14,Lu:14,Santos:15,Adolfo:16,Deffner:16,Song:16,Santos:16,Lutz:18,Funo:17,Baris:19,Chen:21}, with 
many recent experimental realizations~\cite{An:16,Hu:18,Wang:19,Zhang:18,Deng:18}.

In a real physical scenario, where the quantum system is coupled with its surrounding environment, the adiabatic approximation requires a reformulation 
so that it is applicable to a non-unitary evolution. In that case, the closed system adiabatic picture of a decoupled evolution of the Hamiltonian eigenspaces 
with distinct energy eigenvalues is replaced for a decoupled evolution of Lindblad-Jordan eigenspaces with distinct eigenvalues of the Lindbladian superoperator~\cite{Sarandy:05-1} 
(for alternative, but similar reformulations, see Refs.~\cite{Yi:07,Venuti:16}). This notion of adiabaticity has been consistently applied in different scenarios, 
such as quantum computation~\cite{Sarandy:05-2}, eigenstate tracking of open quantum systems~\cite{Jing:16}, and quantum thermodynamics~\cite{Hu:19-d}. 
As a further application, this open system adiabatic approximation has also been used to build a theory of shortcuts to adiabaticity via transitionless evolutions, 
as shown by G. Vacanti~\textit{et al.}~\cite{Vacanti:14}. As an extension of closed system TQD, the open system TQD is established for time-local master equations 
by adding a counterpart to the relevant Lindbladian governing the dynamics. 
In turn, it requires ability of controlling both fields 
and decohering rates, so that environment engineering is taken as a tool to drive the system along an open system adiabatic path. 
Recently, this protocol has been experimentally investigated in circuit quantum electrodynamics~\cite{Yin:21}.

In this work, we generalize the theory of TQD for open systems introduced in Ref.~\cite{Vacanti:14} for the case of shortcuts to adiabaticity 
exhibiting a phase (gauge) freedom. This brings to the realm of open systems the generalized TQD approach for closed systems theoretically 
proposed in Ref.~\cite{Santos:18-b} and experimentally realized in Ref.~\cite{Hu:18}. By considering the adiabatic dynamics in open systems 
and by taking the phases accompanying the evolution as free parameters, we derive a general counter-diabatic Lindbladian implementing 
arbitrary paths in a non-unitary evolution. We then show that, in addition to the path acceleration expected in the TQD dynamics, this set of 
free parameters can considerably simplify the underlying master equations allowing, for example, the derivation of time-independent Lindbladians. 
Moreover, as shown for closed systems, the phase freedom is potentially able to provide smooth energy requirements for local fields and interactions throughout 
the system dynamics~\cite{Santos:18-b}. This comes at the expense of convenient environment engineering, which may be achieved by employing 
suitable quantum control techniques. We then discuss the feasibility of the method through its illustration in the Deutsch quantum algorithm and the Landau-Zener Hamiltonian.

The manuscript is organized as follows. In Sec.~\ref{SecGenCS} we review some elements of the generalized TQD in closed systems, where we highlight the 
usefulness of the generalized phases introduced in Ref.~\cite{Santos:18-b}.  In Sec.~\ref{SecTTQDOS} 
we analyze the results obtained in Ref.~\cite{Vacanti:14}. As a contribution to this previous work, we will rewrite here the counter-diabatic Lindbladian 
introduced by Ref.~\cite{Vacanti:14} in terms of the right and left eigenbases of the Lindbladian superoperator. 
Then, in Sec.~\ref{SecGTQDOS} we discuss how the generalized phases can be employed to implement TQD for time-local master 
equations. In Sec.~\ref{SecApplications}, we then illustrate our results through some applications in quantum control and quantum computation.

\section{Preliminary results} \label{SecGenCS}

\subsection{Transitionless quantum driving in closed systems}

Consider a $D_{\text{S}}$-dimensional Hilbert space $\Hcal_{\text{S}}$ describing a quantum system driven by a time-dependent Hamiltonian 
$H(t)$ acting on $\Hcal_{\text{S}}$, whose instantaneous eigenvectors are $\ket{n(t)}$ and their corresponding energies $E_{n}(t)$. 
Under sufficiently slow evolution~\cite{Born:28,Kato:50,Messiah:Book}, 
the system will follow an adiabatic dynamics governed by the adiabatic evolution operator
\begin{eqnarray}
U_{\text{ad}}(t;t_{0}) = \sum_{n=0}^{D_{\text{S}}-1} e^{-i \int_{t_{0}}^{t} \theta_{n}^{\text{ad}}(\xi) d\xi } \ket{n(t)}\bra{n(t_{0})} \text{ , } \label{AdSol}
\end{eqnarray}
where $\theta_{n}^{\text{ad}}(t)$ is the adiabatic phase that accompanies the evolution of the $n$-th eigenstate, which is given by~\cite{Berry:84}
\begin{eqnarray}
\theta_{n}^{\text{ad}}(t) = \frac{E_{n}(t)}{\hbar} + i \interpro{\dot{n}(t)}{n(t)} \text{ , }
\end{eqnarray}
with the dot symbol  ``$\cdot$" denoting time derivative (this notation is adopted througout the paper). 
We can speed up such evolution through TQD methods~\cite{Demirplak:03,Demirplak:05,Berry:09}. Indeed, we can achieve the dynamics provided by the 
operator $U_{\text{ad}}(t;t_{0})$ at arbitrary finite time~\cite{Santos:15}. This occurs by letting the system evolve under the action of the standard TQD Hamiltonian
\begin{equation}
H_{\text{tqd}}(t) = H(t) + i\hbar\sum_{n=0}^{D_{\text{S}}-1} \left( \ket{\dot{n}(t)}\bra{n(t)} + \interpro{\dot{n}(t)}{n(t)}\ket{n(t)}\bra{n(t)} \right) \text{ , } \label{HtqdCStheory}
\end{equation}
where the first term is the original Hamiltonian, whose spectral decomposition reads $H(t)=\sum_n E_n(t) \ket{n(t)}\bra{n(t)}$, and the second term is the 
so-called counter-diabatic Hamiltonian $H_{\text{cd}}(t)$, whose effect is to inhibit the typical diabatic behavior brought by fast evolutions. 
Particularly, there are a number of situations in which the system is initially prepared in a 
single eigenstate $\ket{\psi(t_{0})}=\ket{k(t_{0})}$ of the Hamiltonian $H(t_{0})$. In these cases, the adiabatic dynamics yields the evolved state 
\begin{eqnarray}
\ket{\psi(t)} = U_{\text{ad}}(t;t_{0})\,\ket{\psi(t_{0})} = e^{-i \int_{t_{0}}^{t} \theta_{k}^{\text{ad}}(\xi) d\xi }\ket{k(t)} \text{ , }
\end{eqnarray}
where the quantum adiabatic phase $\theta_{k}^{\text{ad}}(t)$ works as a global phase. Therefore, it can be neglected in many applications, such as  
the realization of quantum gates. For these cases, it means that any phase that appears during the evolution does not contribute to the TQD evolution. 
It is then possible to derive an alternative TQD protocol with arbitrary quantum phases [not necessarily the adiabatic phase $\theta_{n}^{\text{ad}}(t)$]. 
This introduces a phase freedom and constitutes the \textit{generalized} TQD dynamics~\cite{Santos:18-b}. 
Thus, we can define the generalized TQD evolution operator 
\begin{eqnarray}
U_{\text{tqd}}^{\text{gen}}(t;t_{0}) = \sum_{n=0}^{D_{\text{S}}-1} e^{-i \int_{t_{0}}^{t} \theta_{n}(\xi) d\xi } \ket{n(t)}\bra{n(t_{0})} \text{ , } \label{GenTQDSol}
\end{eqnarray}
where $\{\theta_{n}(t)\}$ is a set of arbitrary phases to be freely adjusted according to the desired dynamics. 
From such operator, we derive the generalized TQD Hamiltonian that drives the system through this path as~\cite{Santos:18-b}
\begin{eqnarray}
H_{\text{tqd}}^{\text{gen}}(t) = i\hbar \sum_{n=0}^{D_{\text{S}}-1} \left( \ket{\dot{n}(t)}\bra{n(t)} - i\theta_{n}(t) \ket{n(t)}\bra{n(t)} \right) \text{ . }
\end{eqnarray}
The potential benefits of this approach have been illustrated in quantum computation and quantum control. Generalized TQD can provide feasible time-independent 
TQD Hamiltonians, which can be applied to implement quantum gates through controlled evolutions and to speed up the dynamics of two-level atomic systems under  
Landau-Zener transitions~\cite{Santos:18-b}. Experimentally, generalized TQD has been used to design energy enhanced TQD microwave fields to implement shortcuts 
to adiabaticity in trapped ion system~\cite{Hu:18} and nuclear magnetic resonance~\cite{Santos:20b}. 

\subsection{Standard transitionless quantum driving in open systems}\label{SecTTQDOS}

We assume a time-local open system dynamics described by a superoperator $\Lcal[\bullet]$, which governs both the unitary (coherent) dynamics and 
the non-unitary contribution (due to the coupling with the reservoir). The master equation underlying the system evolution is provided by $\dot{\rho}(t)\!=\!\Lcal[\rho(t)]$. 
In the context of the open system adiabatic dynamics, it is convenient to rewrite the master equation in the superoperator formalism as~\cite{Sarandy:05-1} (See Appendix~\ref{SecSuperOpForm})
\begin{align}
\dket{\dot{\rho}(t)} = \Lmath(t)\dket{\rho(t)} , \label{LindEq}
\end{align}
where the vector $\dket{\rho(t)}$ and the Lindbladian superoperator $\Lmath(t)$ in above equation are written in a matrix basis composed by 
$(D_{\text{S}}\!\times\!D_{\text{S}})$-dimensional traceless matrices $\sigma_{n}$ satisfying the relation $\trs{\sigma_{n} \sigma_{m}}\!=\! D_{\text{S}}\delta_{nm}$. 
Then, we have that  $\dket{\rho(t)}$ is a $D_{\text{S}}^2$-dimensional ``coherence" vector in Hilbert-Schmidt space~\cite{Muynck:Book}, 
whose components are $\varrho_{n}(t)\!=\!\trs{\rho(t)\sigma_{n}^{\dagger}}$ and a $(D_{\text{S}}^2\times D_{\text{S}}^2)$-dimensional superoperator $\Lmath(t)$ with matrix elements
$\Lmath_{ki}(t)=(1/D_{\text{S}})\trs{\sigma_{k}^{\dagger} \Lcal [ \sigma_{i} ]}$. 
The inner product between two coherence vectors associated with density operators 
$\xi_{1}$ and $\xi_{2}$ is given by $\dinterpro{\xi_{1}}{\xi_{2}} = (1/D_{S})\trs{\xi^{\dagger}_{1}\xi_{2}}$, where the conjugate coherence vector $\dbra{\xi_{1}}$ has 
components given by $\trs{\xi^{\dagger}_{1}\sigma_{n}}$. In particular, for a two-level system, the Hermitian Pauli basis $\Ocal_{\text{tls}} = \{\1, \sigma_{x} , \sigma_{y} , \sigma_{z} \}$ 
is a convenient choice, but we can adopt alternative bases depending on the application~\cite{Santos:21}.  

In the formalism of superoperators, the adiabatic dynamics is defined from the instantaneous decoupled evolution of Jordan blocks of $\Lmath(t)$. 
As shown in Ref.~\cite{Santos:20c}, the adiabatic dynamics is well characterized by the open system evolution operator $\Vcal_{\text{ad}} (t,t_{0}) = \sum_{\beta = 0}^{N-1} \Vcal_{\beta} (t,t_{0})$, 
where each element $\Vcal_{\beta} (t,t_{0})$ reads
\begin{equation}
\Vcal_{\beta} (t,t_{0}) = e^{\int_{t_{0}}^{t} \lambda_{\beta}(\xi)d\xi} \sum_{n_{\beta} = 1}^{N_{\beta}} \sum_{m_{\beta} = 1}^{N_{\beta}} v_{{n_\beta}m_{\beta}}(t)\dket{\Dcal_{\beta}^{n_{\beta}}(t)}\dbra{\Ecal_{\beta}^{m_{\beta}}(t_{0})} \label{EqUalpha}\text{ , }
\end{equation}
with the elements $v_{n_{\beta}m_{\beta}}(t)$ accounting for inner transitions within a single Jordan block~\cite{Santos:20c} and $\lambda_{\beta}(t)$ being the instantaneous eigenvalue associated 
to the $\beta$-th block, whose right (left) quasi-eigenvectors  $\dket{\Dcal_{\beta}^{m_{\beta}}(t)}$ ($\dbra{\Ecal_{\beta}^{m_{\beta}}(t)}$) obey
\begin{subequations}
	\label{EqEqEigenStateL}
	\begin{align}
	\Lmath(t)\dket{\Dcal_{\alpha}^{n_{\alpha}}(t)} &= \dket{\Dcal_{\alpha}^{(n_{\alpha}-1)}(t)} + \lambda_{\alpha}(t)\dket{\Dcal_{\alpha}^{n_{\alpha}}(t)} \text{ , } \label{EqEqEigenStateLa} \\
	\dbra{\Ecal_{\alpha}^{n_{\alpha}}(t)}\Lmath(t) &= \dbra{\Ecal_{n}^{(n_{\alpha}+1)}(t)} + \dbra{\Ecal_{\alpha}^{n_{\alpha}}(t)}\lambda_{\alpha}(t) \text{ , } \label{EqEqEigenStateLb}
	\end{align}
\end{subequations}
with $\dket{\Dcal_{\alpha}^{(0)}\!(t)}$ and $\dbra{\Ecal_{\alpha}^{(N_{\alpha}+1)}\!(t)}$ denoting vanishing vectors. 
The sets $\{\dket{\Dcal_{\alpha}^{n_{\alpha}}(t)}\}$ and $\{\dbra{\Ecal_{\alpha}^{n_{\alpha}}(t)}\}$ satisfy the bi-orthonormalization condition 
$\dinterpro{\Ecal_{m}^{\beta}(t)}{\Dcal_{n}^{\alpha}(t)} = \delta_{mn}\delta_{\beta\alpha}$. 
For one-dimensional Jordan block decomposition of the Lindbladian $\Lmath(t)$, we have block dimension $N_{\alpha}=1 \,\, ,\,\, \forall \alpha$. In this case, we define 
$\dket{\Dcal_{\alpha}^{(1)}(t)} \equiv \dket{\Dcal_{\alpha}(t)}$ and  $\dbra{\Ecal_{\alpha}^{(1)}(t)} \equiv \dbra{\Ecal_{\alpha}(t)}$. Then, 
$\Vcal_{\text{ad}} (t,t_{0})$ becomes
\begin{align}
\Vcal_{\text{ad}}^{\text{1D}}(t,t_{0}) = \sum_{\alpha = 0}^{N-1} e^{\int_{t_{0}}^{t} \Lambda_{\alpha}(\xi)d\xi}\dket{\Dcal_{\alpha}(t)}\dbra{\Ecal_{\alpha}(t_{0})} \text{ . } \label{OpEvo1D}
\end{align}
with $\Lambda_{\alpha}(t)=\lambda_{\alpha}(t)-\dinterpro{\Ecal_{\alpha}(t)}{\dot{\Dcal}_{\alpha}(t)}$ being the generalized adiabatic phase accompanying the dynamics of the $n$-th eigenvector~\cite{Santos:20c}. 

In the same direction as the TQD for closed systems, the TQD can be introduced here to mimic the adiabatic dynamics, 
but in this case the adiabatic behavior is dictated by its generalized version for open systems. 
In Ref.~\cite{Vacanti:14}, G. Vacanti \textit{et al.} have provided an interesting and useful discussion of how we should deal with TQD in such systems. 
Similarly as provided for closed systems, the idea is to define a counter-diabatic term which, when added to the original Lindbladian, 
provides open system TQD at finite time. To this end, the authors defined a transformation $C(t)$ which depends on the set of instantaneous 
right quasi-eigenstate of the Lindbladian. From this approach, it is then possible to find the counter-diabatic Lindbladian $\Lmath_{\text{cd}}(t)$ 
in terms of $C(t)$. By following a generalized path, we can show here how to formulate $\Lmath_{\text{cd}}(t)$ in an analogous notation as that used for 
counter-diabatic Hamiltonians in closed systems~\cite{Berry:09}.

First, let us briefly review the proposal for speeding up an adiabatic dynamics via TQD as originally proposed in Ref.~\cite{Vacanti:14}. 
The idea is based on achieving the perfect decoupling of Jordan-Lindblad eigenspaces, in complete agreement with the definition of adiabaticity. 
To this end, it is considered the similarity transformation $\Lmath_{\text{J}}(t) = C^{-1}(t) \Lmath(t) C(t)$, with $\Lmath_{\text{J}}(t)$ denoting the Jordan 
canonical form of $\Lmath(t)$ and the superoperator $C(t)$ defined by
\begin{eqnarray}
C(t) = \sum_{\mu=0}^{N-1}\sum_{n_{\mu}=1}^{N_{\mu}} \dket{\Dcal_{\mu}^{n_{\mu}}(t)} \dbra{\sigma_{\mu}^{n_{\mu}}} \text{ , }
\label{Csupop}
\end{eqnarray}
where $\{\dbra{\sigma_{\mu}^{n_{\mu}}}\}$ is a set of time-independent vectors corresponding to a basis of traceless orthogonal matrices $\{\sigma_n\}$.  
From this transformation, Eq.~\eqref{LindEq} becomes
\begin{eqnarray}
\left[\Lmath_{\text{J}}(t) + \dot{C}^{-1}(t)C(t)\right] \dket{\rho(t)}_{\text{J}} = \dket{\dot{\rho}(t)}_{\text{J}} \text{ , }
\label{simtrans}
\end{eqnarray}
with $\dket{\rho(t)}_{\text{J}}  = C^{-1}(t) \dket{\rho(t)}$. Then, following Ref.~\cite{Vacanti:14}, we obtain
\begin{eqnarray}
\left[\Lmath_{\text{J}}(t) + \Lmath_{\text{J}}^{\prime}(t) + \Lmath_{\text{nd}}^{\prime}(t)\right] \dket{\rho(t)}_{\text{J}} = \dket{\dot{\rho}(t)}_{\text{J}} \text{ , }
\end{eqnarray}
where the operator $\dot{C}^{-1}(t)C(t)$ has been split into two parts, a block-diagonal operator $\Lmath_{\text{J}}^{\prime}(t)$ 
and a second off-diagonal contribution $\Lmath_{\text{nd}}^{\prime}(t)$ given, respectively, by
\begin{align}
\Lmath_{\text{J}}^{\prime}(t) &= \sum_{\mu=0}^{N-1}\sum_{n_{\mu}=1}^{N_{\mu}} \sum_{l_{\mu}=1}^{N_{\mu}} C_{\mu\mu}^{n_{\mu}l_{\mu}}(t)
\dket{\sigma_{\mu}^{n_{\mu}}}\dbra{\sigma_{\mu}^{l_{\mu}}} \text{ , } \\
\Lmath_{\text{nd}}^{\prime}(t) &= \sum_{\nu\neq\mu}^{N-1}\sum_{n_{\nu}=1}^{N_{\nu}}\sum_{\mu=0}^{N-1}\sum_{n_{\mu}=1}^{N_{\mu}}
C_{\nu\mu}^{n_{\nu}n_{\mu}}(t)\dket{\sigma_{\nu}^{n_{\nu}}}\dbra{\sigma_{\mu}^{n_{\mu}}} \text{ , }
\end{align}
with coefficients $C_{\nu\mu}^{n_{\nu}n_{\mu}}(t) = \dbra{\sigma_{\nu}^{n_{\nu}}} \dot{C}^{-1}(t)C(t)\dket{\sigma_{\mu}^{n_{\mu}}}$. 
The operator $\Lmath_{\text{nd}}^{\prime}(t)$ is associated with transition between Jordan blocks. In order to inhibit the effects of $\Lmath_{\text{nd}}^{\prime}(t)$, 
we introduce an additional operator $\Lmath_{\text{cd}}(t)$ given by
\begin{align}
\Lmath_{\text{cd}}(t) &= - C(t)\Lmath_{\text{nd}}^{\prime}(t)C^{-1}(t) \text{ . }
\label{sLCD}
\end{align}
Thus, the \textit{standard} TQD Lindbladian reads
\begin{eqnarray}
\Lmath_{\text{Stqd}}(t) = \Lmath(t) + \Lmath_{\text{cd}}(t) \text{ , } \label{EqTradTQDL}
\end{eqnarray}
This superoperator has been obtained in Ref.~\cite{Vacanti:14} as the \textit{counter-diabatic Lindbladian} to implement TQD in open systems. 
As a further step, we can go beyond Ref.~\cite{Vacanti:14} at this point and rewrite Eq.~(\ref{sLCD}) in a more convenient way as (see Appendix~\ref{ApVacantiLCD})
\begin{align}
\Lmath_{\text{cd}}(t) = \sum_{\mu=0}^{N-1}\sum_{n_{\mu}=1}^{N_{\mu}} &\left[\dket{\dot{\Dcal}_{\mu}^{n_{\mu}}} \dbra{\Ecal_{\mu}^{n_{\mu}}} - \sum_{k_{\mu}=1}^{N_{\mu}}\Gcal_{\mu\mu}^{n_{\mu}k_{\mu}} \dket{\Dcal_{\mu}^{n_{\mu}}}\dbra{\Ecal_{\mu}^{k_{\mu}}} \right] \text{ . }
\label{firstR}
\end{align}
where $\Gcal_{\mu\mu}^{n_{\mu}k_{\mu}}(t)=\dinterpro{\Ecal_{\mu}^{n_{\mu}}(t)}{\dot{\Dcal}_{\mu}^{k_{\mu}}(t)}$.
Notice the formally identical structure for $\Lmath_{\text{cd}}(t)$ in comparison with the standard counter-diabatic Hamiltonian $H_{\text{cd}}(t)$ in Eq.~\eqref{HtqdCStheory}. 
Eq.~(\ref{firstR}) is a first contribution of our work. The dynamics induced by the Lindbladian $\Lmath_{\text{Stqd}}(t)$ will be here referred to as 
the {\it standard} TQD evolution, since it allows for the exact mimic of the adiabatic path in open systems (see Appendix~\ref{ApVacantiLCD}). 
In particular, for a one-dimensional Jordan block decomposition, we have
\begin{align}
\Lmath_{\text{cd}}^{\text{1D}}(t) &= \sum_{\alpha=0}^{N-1} \dket{\dot{\Dcal}_{\alpha}(t)} \dbra{\Ecal_{\alpha}(t)} -\Gcal_{\alpha}(t) \dket{\Dcal_{\alpha}(t)}\dbra{\Ecal_{\alpha}(t)} \text{ , }
\end{align}
with $\Gcal_{\alpha}(t)=\dinterpro{\Ecal_{\alpha}(t)}{\dot{\Dcal}_{\alpha}(t)}$, so that
\begin{eqnarray}
\Lmath_{\text{Stqd}}^{\text{1D}}(t) = \Lmath(t) + \Lmath_{\text{cd}}^{\text{1D}}(t) 
\end{eqnarray}
is the one-dimensional standard TQD Lindibladian.

\section{Generalized TQD in open systems}\label{SecGTQDOS}

In closed quantum systems, we can design a Hamiltonian $H(t)$ that implements a target unitary evolution by 
the method of  inverse engineering (see, e.g., Refs.~\cite{Kang:16,Santos:18-a}). This can be achieved by 
using $H(t) = i\hbar \dot{U}(t)U^{\dagger}(t)$, with $U(t)$ denoting the evolution operator. Equivalently, we can 
show that we can perform inverse engineering for Lindbladian superoperators in non-unitary dynamics. 
To this end, we assume open systems driven by invertible dynamical maps, with the evolved density 
operator $\dket{\rho(t)}$ provided by a non-unitary superoperator $\Vcal \left( t,t_{0}\right)$ as 
$\dket{\rho(t)} = \Vcal \left( t,t_{0}\right) \dket{\rho(t_{0})}$. We can then introduce an inversely engineered Lindbladian 
$\Lmath_{\text{ie}}(t)$ implementing the dynamics induced by $\Vcal \left( t,t_{0}\right)$ by taking (see Appendix~\ref{ApenIEOS})
\begin{eqnarray}
\Lmath_{\text{ie}}(t) = \dot{\Vcal} \left( t,t_{0}\right)\Vcal^{-1} \left( t,t_{0}\right) \text{ . } \label{Linver}
\end{eqnarray}
Since the evolution is encoded in the superoperator $\Vcal \left( t,t_{0}\right)$, TQD can be directly approached by this strategy. In particular, 
by using an engineered Lindbladian $\Lmath_{\text{ie}}(t)$ from Eq.~\eqref{Linver}, we could drive the system through an adiabatic path so 
that the transitions among Jordan blocks are suppressed~\cite{Santos:20c}. Furthermore, we can design $\Lmath_{\text{ie}}(t)$ 
capable of implementing shortcuts to the non-unitary adiabatic evolution through general TQD protocols.

\subsection{Generalized TQD for one-dimensional Jordan block}

Let us consider the case of one-dimensional Jordan decomposition for the original Lindbladian $\Lmath(t)$. By requiring the decoupling of the 
Jordan blocks and by allowing arbitrary phases throughout the evolution, we write a generalized TQD evolution operator as
\begin{eqnarray}
\Vcal_{\text{Gtqd}}^{\text{1D}}(t,t_{0}) = \sum_{\alpha = 0}^{N-1} e^{\int_{t_{0}}^{t} \Theta_{\alpha}(\xi)d\xi} \dket{\Dcal_{\alpha}(t)}\dbra{\Ecal_{\alpha}(t_{0})} \text{ . }
\end{eqnarray}
where the function $\Theta_{\alpha}(t) \in \Cmath$ is a free generic phase. From Eq.~\eqref{Linver}, we can then show that the generalized TQD Lindbladian for 
one-dimensional Jordan decomposition is given by (See Appendix~\ref{ApRecTTQDL})
\begin{eqnarray}
\Lmath_{\text{Gtqd}}^{\text{1D}}(t) = \sum_{\alpha = 0}^{N-1}\left( \dket{\dot{\Dcal}_{\alpha}(t)} \dbra{\Ecal_{\alpha}(t)} + \Theta_{\alpha}(t) \dket{\Dcal_{\alpha}(t)}\dbra{\Ecal_{\alpha}(t)} \right) . \,\,\,\,\,\, \label{GenL1D}
\end{eqnarray}
As expected, we can choose $\Theta_{\alpha}(t)$ as to recover the adiabatic phase. However, by letting $\Theta_{\alpha}(t)$ as a free parameter, we can achieve an infinite family of TQD evolutions mimicking the 
adiabatic dynamics up to a quantum phase. As we will show, this phase freedom may simplify the physical implementation.    
Notice also that the set of parameters $\{\Theta_{n}(t)\}$ cannot be considered completely arbitrary because we usually do not prepare the system in a particular eigenstate of $\Lmath(t)$ at the beginning of the evolution. 
Such result will be illustrated in Sec.~\ref{SecApplications}. Nonetheless, we can already provide some advantages of the generalized phases concerning the feasibility of the operator $\Lmath_{\text{Gtqd}}^{\text{1D}}(t)$. 
Indeed, it is possible to derive a theorem that tells us about the situations for which we could obtain a time-independent TQD Lindbladian.

\begin{theorem}\label{TheoTimeIndL}
	Let $\Lmath(t)$ be a Lindblad superoperator that admits one-dimensional Jordan block decomposition, with non-crossing eigenvalues $\lambda_{\alpha}(t)$ associated with right- and left-eigenvectors 
	$\dket{\Dcal_{\alpha}(t)}$ and $\dbra{\Ecal_{\alpha}(t)}$, respectively. If the sets $\{\dket{\Dcal_{\alpha}(t)}\}$ and $\{\dbra{\Ecal_{\alpha}(t)}\}$ obey
	\begin{eqnarray}
	\dinterpro{\Ecal_{\eta}(t)}{\dot{\Dcal}_{\beta}(t)} = \dinterpro{\Ecal_{\eta}(t_{0})}{\dot{\Dcal}_{\beta}(t_{0})} e^{\int_{t_{0}}^{t} \left[ \Gcal_{\eta}(\xi) - \Gcal_{\beta}(\xi) \right] d\xi } \text{ , } \label{EqTheo1}
	\end{eqnarray}
	for every $\eta$ and $\beta$, then we can derive a {\it time-independent} TQD Lindblad superoperator $\Lmath_{\textnormal{Gtqd}}^{1\textnormal{D}}$ by adopting generalized phases given by
	\begin{eqnarray}
	\overline{\Theta}_{\eta}(t) = - \dinterpro{\Ecal_{\eta}(t)}{\dot{\Dcal}_{\eta}(t)} \text{ , } \label{EqTheo2}
	\end{eqnarray}
	for every $\eta$.
\end{theorem}
The proof is provided in Appendix~\ref{ApTheoTimeIndL}. Such theorem is potentially useful when we have limited ability of implementing (or simulating) reservoirs effects and/or 
when we do not have optimal control on the time-dependent external parameters that act on the system. 
As a by-product, if the sets $\{\dket{\Dcal_{\alpha}(t)}\}$ and $\{\dbra{\Ecal_{\alpha}(t)}\}$ obey the particular condition
\begin{eqnarray}
\frac{d}{dt} \left[\dinterpro{\Ecal_{\eta}(t)}{\dot{\Dcal}_{\beta}(t)}\right] = 0 \text{ , } \label{EqTheo3}
\end{eqnarray}
for every $\eta$ and $\beta$, then a time-independent Lindbladian can be set by choosing $\overline{\Theta}_{\eta}(t)$ such that
\begin{eqnarray}
\overline{\Theta}_{\eta} = \overline{\Theta}_{\beta} = \text{constant} , \,\, \forall \eta,\beta \text{ . } \label{EqTheo4}
\end{eqnarray}
The proof is also provided in Appendix~\ref{ApTheoTimeIndL}. Concerning this last result, we notice that, whenever we have a time-independent eigenvector $\dket{\Dcal_{\nu}}$, 
the corresponding generalized phase $\Theta_{\nu}$ that provides a time-independent Lindbladian does not depend on the rest of the parameters $\Theta_{\beta}(t)$, with $\beta \ne \nu$. 
This means that the choice of $\Theta_{\nu}$ can be made independently from the remaining sectors for the case of a time-independent eigenvector $\dket{\Dcal_{\nu}}$.

\subsection{Generalized TQD for multi-dimensional Jordan blocks}

In order to implement the generalized TQD approach for multi-dimensional Jordan blocks, we start by introducing the evolution operator
\begin{eqnarray}
\Vcal_{\text{Gtqd}}(t,t_{0}) = \sum_{\alpha = 0}^{N-1} \sum _{n_{\alpha} = 1}^{N_{\alpha}} \sum _{m_{\alpha} = 1}^{N_{\alpha}} q_{\alpha}^{n_{\alpha}m_{\alpha}}(t)\dket{\Dcal_{\alpha}^{n_{\alpha}}(t)}\dbra{\Ecal_{\alpha}^{m_{\alpha}}(t_{0})} \text{ , } \label{EqUGTQD}
\end{eqnarray}
where the coefficients $q_{\alpha}^{n_{\alpha}m_{\alpha}}(t)$ can be suitably adjusted throughout the dynamics. From Eq.~(\ref{EqUGTQD}), we can see that $\Vcal_{\text{Gtqd}}(t,t_{0})$ drives the system under a transitionless path, but it does not necessarily mimic the exact adiabatic solution. The generalized TQD Lindbladian $\Lmath_{\text{Gtqd}}(t)$ then reads 
\begin{eqnarray}
\Lmath_{\text{Gtqd}}(t) = \dot{\Vcal}_{\text{Gtqd}} \left( t,t_{0}\right)\Vcal_{\text{Gtqd}}^{-1} \left( t,t_{0}\right) \text{ . }
\label{EqLGTQDfromU}
\end{eqnarray}
By using the Eq.~\eqref{EqUGTQD} in Eq.~(\ref{EqLGTQDfromU}), we obtain (see Appendix~\ref{ApGeneralizedLCD})
\begin{align}
\Lmath_{\text{Gtqd}}(t,t_{0}) &= \sum_{\alpha = 0}^{N-1} \sum _{n_{\alpha} = 1}^{N_{\alpha}} 
\sum _{k_{\alpha} = 1}^{N_{\alpha}} \sum _{l_{\alpha} = 1}^{N_{\alpha}} 
\dot{q}_{\alpha}^{n_{\alpha}k_{\alpha}}(t)\tilde{q}_{\alpha}^{k_{\alpha}l_{\alpha}}(t)\dket{\Dcal_{\alpha}^{n_{\alpha}}(t)}\dbra{\Ecal_{\alpha}^{l_{\alpha}}(t)} \nonumber \\
&+ \sum_{\alpha = 0}^{N-1} \sum _{n_{\alpha} = 1}^{N_{\alpha}}\dket{\dot{\Dcal}_{\alpha}^{n_{\alpha}}(t)}\dbra{\Ecal_{\alpha}^{n_{\alpha}}(t)}
\label{LG_tqd}
\text{ , }
\end{align}
where the coefficients $\tilde{q}_{\alpha}^{k_{\alpha}l_{\alpha}}(t)$ are associated with the operator $\Vcal_{\text{Gtqd}}^{-1}(t,t_{0})$, which satisfies $\Vcal_{\text{Gtqd}}(t,t_{0})\Vcal_{\text{Gtqd}}^{-1}(t,t_{0})=\1$. 
It is also possible to prove that the coefficients $\tilde{q}_{\alpha}^{k_{\alpha}l_{\alpha}}$ satisfies (See Appendix~\ref{ApGeneralizedLCD})
\begin{eqnarray}
\sum _{m_{\kappa} = 1}^{N_{\kappa}}
q_{\kappa}^{l_{\kappa}m_{\kappa}} \tilde{q}_{\kappa}^{m_{\kappa}i_{\kappa}} = \delta_{l_{\kappa}i_{\kappa}} \label{EqqDelta}\text{ . }
\end{eqnarray}
The TQD Lindibladian $\Lmath_{\text{Gtqd}}(t,t_{0})$ generalizes the results presented in Sec.~\ref{SecGenCS}. While in standard case the feasibility of the TQD Lindbladian 
depends on the spectrum of the original Lindbladian $\Lmath(t,t_{0})$, in the generalized approach the implementation of $\Lmath_{\text{Gtqd}}(t,t_{0})$ depends on both the 
spectrum of $\Lmath(t,t_{0})$ and the free parameters $q_{\alpha}^{n_{\alpha}m_{\alpha}}(t)$ and $\tilde{q}_{\alpha}^{k_{\alpha}l_{\alpha}}(t)$. Those parameters can be used 
to optimize a TQD evolution provided the experimental setup available.

Let us analyze now the particular case $q_{\alpha}^{n_{\alpha}m_{\alpha}}(t) = e^{\int_{t_{0}}^{t} \lambda_{\alpha}(\xi)d\xi}v_{n_{\alpha}m_{\alpha}}(t)$, such as in Eq.~\eqref{EqUalpha}. 
Then, we have ${\Vcal}_{\text{Gtqd}} \left( t,t_{0}\right) = {\Vcal}_{\text{ad}} \left( t,t_{0}\right)$, so that the standard TQD Lindbladian in Eq.~\eqref{EqTradTQDL} is expected to be recovered. 
Indeed, by using the adiabatic choice for  $q_{\alpha}^{n_{\alpha}m_{\alpha}}(t) $, we obtain the generalized TQD Lindbladian (see Appendix~\ref{ApRecTLindGenBlock})
\begin{align}
\Lmath_{\text{Gtqd}}^{q=v}(t)
& = \sum_{\alpha = 0}^{N-1} \sum _{n_{\alpha} = 1}^{N_{\alpha}}
\lambda_{\alpha}(t)
\dket{\Dcal_{\alpha}^{n_{\alpha}}(t)}\dbra{\Ecal_{\alpha}^{n_{\alpha}}(t)} + \dket{\dot{\Dcal}_{\alpha}^{n_{\alpha}}(t)}\dbra{\Ecal_{\alpha}^{n_{\alpha}}(t)}\nonumber \\
& +\sum_{\alpha = 0}^{N-1} \sum _{n_{\alpha} = 1}^{N_{\alpha}} \sum _{k_{\alpha} = 1}^{N_{\alpha}} \sum _{l_{\alpha} = 1}^{N_{\alpha}} 
\dot{v}_{n_{\alpha}k_{\alpha}}(t) \tilde{v}_{k_{\alpha}l_{\alpha}}(t)\dket{\Dcal_{\alpha}^{n_{\alpha}}(t)}\dbra{\Ecal_{\alpha}^{l_{\alpha}}(t)} \text{ , } \label{GenTQDOS}
\end{align}
where we used the existence of the inverse evolution superoperator $\Vcal_{\text{ad}}^{-1} (t,t_{0})$, 
such that $\Vcal_{\text{ad}}  (t,t_{0})\Vcal_{\text{ad}} ^{-1} (t,t_{0})\!=\!\1$. Explicitly, we write $\Vcal_{\text{ad}}^{-1} (t,t_{0}) = \sum_{\alpha = 0}^{N-1} \Vcal_{\alpha}^{-1} (t,t_{0})$, where~\cite{Santos:20c}
\begin{align}
\Vcal_{\alpha}^{-1} (t,t_{0}) = e^{-\int_{t_{0}}^{t} \lambda_{\alpha}(\xi)d\xi} \sum _{n_{\alpha} = 1}^{N_{\alpha}} \sum _{m_{\alpha} = 1}^{N_{\alpha}} \tilde{v}_{n_{\alpha}m_{\alpha}}(t)\dket{\Dcal_{\alpha}^{n_{\alpha}}(t_{0})}\dbra{\Ecal_{\alpha}^{m_{\alpha}}(t)} , \label{UinverGenBlock}
\end{align}
where the coefficients $\tilde{v}_{n_{\alpha}m_{\alpha}}(t)$ and $v_{n_{\beta}m_{\beta}}(t)$ satisfy the relation
\begin{align}
\sum _{j_{\nu} = 1}^{N_{\nu}} 
v_{\ell_{\nu}j_{\nu}}(t)\tilde{v}_{j_{\nu}m_{\nu}}(t) &= \sum _{j_{\nu} = 1}^{N_{\nu}} 
\tilde{v}_{\ell_{\nu}j_{\nu}}(t){v}_{j_{\nu}m_{\nu}}(t) = \delta_{\ell_{\nu}m_{\nu}} \text{ . } \label{EqUcoeffINV}
\end{align}
Eq.~(\ref{EqUcoeffINV}) comes from fact that the operator $\Vcal_{\text{ad}} (t,t_{0})$ is identified as the superoperator that ``block-diagonalizes'' the Lindbladian.
Although we cannot write a spectral decomposition to $\Lmath_{\text{Gtqd}}^{q=v}(t)$, since it is in general non-diagonalizable, it is possible to write a ``quasi"-spectral decomposition for an arbitrary $\Lmath(t)$ as
\begin{align}
\Lmath(t) = \sum_{\alpha = 0}^{N-1}\sum _{n_{\alpha} = 1}^{N_{\alpha}} \left(\dket{\Dcal_{\alpha}^{(n_{\alpha}-1)}(t)}\dbra{\Ecal_{\alpha}^{n_{\alpha}}(t)} + \lambda_{\alpha}(t) \dket{\Dcal_{\alpha}^{n_{\alpha}}(t)}\dbra{\Ecal_{\alpha}^{n_{\alpha}}(t)} \right),
\end{align}
whence it is possible to verify the ``quasi"-eigenvalue equations given by Eq.~\eqref{EqEqEigenStateL}. Since $\Lmath_{\text{Gtqd}}^{q=v}(t)$ has to exactly mimic the adiabatic dynamics, the functions 
$v_{n_{\alpha}k_{\alpha}}(t)$ and $\tilde{v}_{k_{\alpha}l_{\alpha}}(t)$ must obey (see Appendix~\ref{ApRecTLindGenBlock})
\begin{align}
\sum _{k_{\alpha} = 1}^{N_{\alpha}} \dot{v}_{n_{\alpha}k_{\alpha}}(t) \tilde{v}_{k_{\alpha}l_{\alpha}}(t) &= \delta_{n_{\alpha}(l_{\alpha}-1)} - \dinterpro{\Ecal_{\alpha}^{n_{\alpha}}(t)}{\dot{\Dcal}_{\alpha}^{l_{\alpha}}(t)} \text{ . }
\end{align}
Therefore, Eq.~\eqref{GenTQDOS} yields
\begin{eqnarray}
\Lmath_{\text{Gtqd}}^{q=v}(t) = \Lmath(t) + \Lmath_{\text{cd}}(t) = \Lmath_{\text{Stqd}}(t) \text{ , }
\end{eqnarray}
which shows how the generalized version of TQD can recover the standard TQD as provided in Ref.~\cite{Vacanti:14}.

\section{Applications} \label{SecApplications}

\subsection{Deutsch algorithm under dephasing}

Our first application is the Deutsch algorithm under dephasing. The problem addressed in Deutsch's algorithm~\cite{Deutsch:85} is 
to determinate whether a dichotomic real function $f: x\in\{0,1\} \rightarrow f(x)\in\{0,1\}$ is either \textit{constant} (the output $f(x)$ 
is the same regardless the input value $x$) or \textit{balanced} (the output $f(x)$ assumes different values according with the input value $x$). 
We denote $\Ocal_{f}$ as the oracle operator, which is capable to compute $f$. The oracle $\Ocal_{f}$ is given by~\cite{Sarandy:05-2}
\begin{eqnarray}
\Ocal_{f} = (-1)^{f(0)}\ket{0}\bra{0} + (-1)^{f(1)}\ket{1}\bra{1} \text{ . }
\end{eqnarray}
Thus, we can write the Hamiltonian that implements the adiabatic solution for the problem as
\begin{eqnarray}
H^{\text{DA}}(t) = U_{f}(t) H_{0} U_{f}^{\dagger}(t) \text{ , }
\end{eqnarray}
where $H_{0} = -\hbar \omega \sigma_{x}$ and $U_{f}(t) = \exp(i\frac{\pi}{2}\frac{t}{\tau}\Ocal_{f})$, with $0 \le t \le \tau$. 
At $t=0$, we have $H^{\text{DA}}(0) =H_{0}$, so that the initial input ground state is written as $\ket{\psi_{\text{inp}}} = \ket{+} = (1/\sqrt{2})(\ket{0}+\ket{1})$. 
When the evolution is slow enough, we can write the output state as the ground state of $H^{\text{DA}}(t)$, which reads
\begin{eqnarray}
\rho_{\text{cs}}^{\text{DA}}(t) = \frac{1}{2} \left[ \1 + g_{\text{c}}(t) \sigma_{x} - g_{\text{s}}(t) \sigma_{y} \right] \text{ , } \label{EqStateDAOp}
\end{eqnarray}
where $g_{\text{c}}(t) = \cos \left( \pi F t/2\tau \right)$ and $g_{\text{s}}(t) = \sin \left( \pi F t/2\tau \right)$, with $F = 1-(-1)^{f(0)+f(1)}$ and the subscript ``cs" denoting 
that $\rho_{\text{cs}}^{\text{DA}}(t)$ is obtained from the adiabatic solution for closed systems.

\subsubsection{Deutsch adiabatic dynamics}

Let us consider an open system dynamics governed by Markovian phase damping, with rate $\gamma(t)$. The evolution is driven by a Lindblad master equation, which is 
given by 
\begin{eqnarray}
\dot{\rho}(t) = - \frac{i}{\hbar} [H^{\text{DA}}(t),\rho(t)] + \gamma(t) \left[ \sigma_{z} \rho(t) \sigma_{z} - \rho(t) \right] \text{ . } \label{EqLindDephDA}
\end{eqnarray}
Let us now rewrite Eq.~(\ref{EqLindDephDA}) in the superoperator formalism, yielding
\begin{eqnarray}
\dket{\dot{\rho}(t)} = \Lmath^{\text{DA}}(t) \dket{\rho(t)} \text{ , }
\end{eqnarray}
where
		\begin{align}
		\Lmath^{\text{DA}}(t) = \begin{bmatrix}
		0 & 0 & 0 & 0 \\ 0 & -2 \gamma(t) & 0 & 2\omega g_{\text{s}}(t) \\ 0 & 0 & -2 \gamma(t) & 2\omega g_{\text{c}}(t) \\ 0 & - 2\omega g_{\text{s}}(t) & - 2\omega g_{\text{c}}(t) & 0
		\end{bmatrix} \text{ . }
		\end{align}
		The right eigenvectors of $\Lmath^{\text{DA}}(t)$ are (the superscript ``t" denotes transpose)
		\begin{subequations}
			\label{EqDARightEigenVec}
			\begin{align}
			\dket{\Dcal^{\text{DA}}_{0}(t)} &= \begin{bmatrix} \text{ } 1 & 0 & 0 & 0\text{ } \end{bmatrix}^{\text{t}} , \\
			\dket{\Dcal^{\text{DA}}_{1}(t)} &= \begin{bmatrix} \text{ } 0 & -g_{\text{c}}(t) & g_{\text{s}}(t) & 0\text{ } \end{bmatrix}^{\text{t}} , \\
			\dket{\Dcal^{\text{DA}}_{2}(t)} &= \begin{bmatrix} 0 & -2\omega g_{\text{s}}(t)/\Delta_{+}(t) & -2\omega g_{\text{c}}(t)/\Delta_{+}(t) & 1 \end{bmatrix}^{\text{t}} , \\
			\dket{\Dcal^{\text{DA}}_{3}(t)} &= \begin{bmatrix}  0 &  -2\omega g_{\text{s}}(t)/\Delta_{-}(t) & -2\omega g_{\text{c}}(t)/\Delta_{-}(t) & 1 \end{bmatrix}^{\text{t}} ,
			\end{align}
		\end{subequations}
		and the left eigenvectors are
		\begin{subequations}
			\label{EqDALeftEigenVec}
			\begin{align}
			\dbra{\Ecal^{\text{DA}}_{0}(t)} &= \begin{bmatrix} \frac{}{} 1 & 0 & 0 & 0\text{ } \end{bmatrix} \text{ , } \\
			\dbra{\Ecal^{\text{DA}}_{1}(t)} &= \begin{bmatrix} \text{ } 0 & -g_{\text{c}}(t) & g_{\text{s}}(t) & 0\text{ } \end{bmatrix} \text{ , } \\
			\dbra{\Ecal^{\text{DA}}_{2}(t)} &= \frac{1}{\Delta(t)} \begin{bmatrix} 0 & \omega g_{\text{s}}(t) & \omega g_{\text{c}}(t) &  \Delta_{+}(t)/2 \end{bmatrix} , \\
			\dbra{\Ecal^{\text{DA}}_{3}(t)} &= \frac{1}{\Delta(t)}\begin{bmatrix}
			\text{ } 0 & - \omega g_{\text{s}}(t) & -\omega g_{\text{c}}(t) & -\Delta_{-}(t)/2\text{ }
			\end{bmatrix} \text{ , }
			\end{align}
		\end{subequations}
with eigenvalues $\lambda_{0}(t) = 0$, $\lambda_{1}(t) = -2 \gamma(t) $, $\lambda_{2}(t) = \Delta_{-}(t) $ and $\lambda_{3}(t) = \Delta_{+}(t)$, where $\Delta_{\pm}(t) = -\gamma(t) \pm \Delta(t)$ and $\Delta(t)\!=\!\sqrt{\gamma^2(t) - 4\omega^2}$.
The non-degenerate spectrum of $\Lmath^{\text{DA}}(t)$ shows that $\Lmath^{\text{DA}}(t)$ can be written in the one-dimensional Jordan block form. By writing the density matrix for the initial state as $\rho^{\text{DA}}(0) = \ket{\psi_{\text{inp}}}\bra{\psi_{\text{inp}}} = \ket{+}\bra{+} = (1/2)(\1+\sigma_{x})$, we can show that the initial state is a linear combination of the vectors $\dket{\Dcal^{\text{DA}}_{0}(0)}$ and $\dket{\Dcal^{\text{DA}}_{1}(0)}$, yielding
\begin{align}
\dket{\rho^{\text{DA}}(0)} &= \begin{bmatrix} \text{ } 1 & 1 & 0 & 0\text{ } \end{bmatrix}^{\text{t}} = \dket{\Dcal^{\text{DA}}_{0}(0)} - \dket{\Dcal^{\text{DA}}_{1}(0)} \text{ . } \label{EqIniStaDA}
\end{align}
Hence, if we let the system evolve adiabatically, we can write the evolved state as~\cite{Santos:20c}
\begin{align}
\dket{\rho_{\text{ad}}^{\text{DA}}(t)} &= \dket{\Dcal^{\text{DA}}_{0}(t)} - e^{-2\int_{t_{0}}^{t} \gamma(\xi)d\xi}\dket{\Dcal^{\text{DA}}_{1}(t)} \label{EqDAAdEvolution} \text{ , }
\end{align}
where we used the adiabatic phase $\Lambda_{1}(t)\!=\!\lambda_{1}(t)\!=\!-2 \gamma(t)$. Now, by rewriting Eq.~(\ref{EqDAAdEvolution}) as a vector in the superoperator formalism, we get
\begin{align}
\dket{\rho_{\text{ad}}^{\text{DA}}(t)} &= \begin{bmatrix}
\text{ } 1 & e^{-2\int_{t_{0}}^{t} \gamma(\xi)d\xi}g_{\text{c}}(t) & - e^{-2\int_{t_{0}}^{t} \gamma(\xi)d\xi}g_{\text{s}}(t) & 0\text{ } 
\end{bmatrix}^{\text{t}} ,
\label{EqAdSolDA}
\end{align}
Then, by determining the coherence vector associated with the density matrix $\rho^{\text{DA}}(t)$ and expressing the result in the Pauli basis, we obtain
\begin{align}
\rho_{\text{ad}}^{\text{DA}}(t) &= \frac{1}{2} \left[ \1 + e^{-2\int_{t_{0}}^{t} \gamma(\xi)d\xi}g_{\text{c}}(t) \sigma_{x} - e^{-2\int_{t_{0}}^{t} \gamma(\xi)d\xi}g_{\text{s}}(t)\sigma_{y} \right] , \label{EqDAAdSolOS}
\end{align}
In the limit $\gamma(t) \rightarrow 0$, we recover the density matrix for the unitary dynamics shown in Eq.~\eqref{EqStateDAOp}, where the output state reads (at $t=\tau$)~\cite{Sarandy:05-2}
\begin{align}
\lim_{\gamma(t) \rightarrow 0}\rho_{\text{ad}}^{\text{DA}}(\tau) &= \rho^{\text{DA}}_{\text{cs}}(\tau) = \frac{1}{2} \left[ \1 + (-1)^{f(0)+f(1)} \sigma_{x} \right]  \text{ , } \label{EqOutStateDAOp}
\end{align}
where we have used $\cos ( \pi F/2) = (-1)^{f(0)+f(1)}$ for $f(x)\in\{0,1\}$, since $F = 1-(-1)^{f(0)+f(1)}$. This solution is the output for an optimal (decoherence-free) situation. 
The experimental implementation of the adiabatic Deutsch algorithm under phase damping has been implemented in trapped ions~\cite{Hu:19-a}, where the adiabatic behavior is achieved for long total evolution time.

\subsubsection{Deutsch standard TQD}

As a first application, let us derive a shortcut to the adiabatic Deutsch algorithm in open systems. 
To begin with, we observe that Eqs.~\eqref{EqDARightEigenVec} and~\eqref{EqDALeftEigenVec} implies in 
$\dinterpro{\Ecal^{\text{DA}}_{\alpha}(t)}{\dot{\Dcal}^{\text{DA}}_{\alpha}(t)} = 0$ for $\alpha\!=\!\{0,1\}$. Therefore, the adiabatic 
phases $\Lambda_{\alpha}(t)=\lambda_{\alpha}(t)-\dinterpro{\Ecal_{\alpha}(t)}{\dot{\Dcal}_{\alpha}(t)}$ for each 
eigenvector are simply given by its corresponding eigenvalue. Then, we get
\begin{subequations}
	\label{EqAdPhaDAOS}
\begin{align}
\Lambda_{0}(t) &= 0 , \quad \Lambda_{1}(t) = -2 \gamma(t), \\ 
\Lambda_{2}(t) &= \lambda_{2}(t) - \frac{\Delta_{-}(t) \dot{\gamma}(t)}{2 \Delta^2(t)}, ~
\Lambda_{3}(t) = \lambda_{3}(t) + \frac{2\omega^2 \dot{\gamma}(t)}{\Delta^2(t)\Delta_{-}(t)}.  \end{align}
\end{subequations}
In particular, we can see that if $\dot{\gamma}(t)\!=\!0$ then $\dinterpro{\Ecal^{\text{DA}}_{\alpha}(t)}{\dot{\Dcal}^{\text{DA}}_{\alpha}(t)}\!=\!0$, for any $\alpha$. This implies in a 
generalized \textit{parallel transport condition} for the adiabatic dynamics in open systems, which is formally analogous to the parallel transport condition for unitary dynamics.

By applying the standard approach for TQD, the counter-diabatic Lindblad superoperator to be added to the adiabatic counterpart reads 
\begin{align}
\Lmath^{\text{DA}}_{\text{cd}}(t) &= -\sum_{\mu=0}^{3} \dinterpro{\Ecal_{\mu}^{\text{DA}}(t)}{\dot{\Dcal}_{\mu}^{\text{DA}}(t)} \dket{\Dcal_{\mu}^{\text{DA}}(t)}\dbra{\Ecal_{\mu}^{\text{DA}}(t)} \nonumber \\
&+ \sum_{\mu=0}^{3} \dket{\dot{\Dcal}_{\mu}^{\text{DA}}(t)} \dbra{\Ecal_{\mu}^{\text{DA}}(t)}\text{ , }
\end{align}
where $\dket{\Dcal_{\mu}^{\text{DA}}(t)}$ and $\dbra{\Ecal_{\mu}^{\text{DA}}(t)}$ are given by Eqs.~\eqref{EqDARightEigenVec} and~\eqref{EqDALeftEigenVec}, respectively. By using the Eqs.~\eqref{EqDARightEigenVec} and Eqs.~\eqref{EqDALeftEigenVec}, we get a counter-diabatic Lindbladian superoperator given by
\begin{align}
\Lmath^{\text{DA}}_{\text{cd}} &= \begin{bmatrix}
	0 & 0 & 0 & 0 \\
	0 & 0 & \frac{F\pi}{2 \tau} & h(t)g_{\text{s}}(t) \\
	0 & -\frac{F\pi}{2 \tau} & 0 & h(t)g_{\text{c}}(t) \\
	0 & h(t)g_{\text{s}}(t) & h(t)g_{\text{c}}(t) & 0 
\end{bmatrix} \text{ . } \label{EqLcdDaUnSimp}
\end{align}
with $h(t) = \omega \dot{\gamma}(t)/\tau[4\omega^2-\gamma^2(t)]$. 
By restricting to the simple case of the parallel transport condition $\dot{\gamma}(t)\!=\!0$, we have $h(t)=0$. Then, Eq.~\eqref{EqLcdDaUnSimp} becomes
\begin{align}
\Lmath^{\text{DA}}_{\text{cd}} = \begin{bmatrix}
0 & 0 & 0 & 0 \\
0 & 0 & \frac{F\pi}{2 \tau} & 0 \\
0 & -\frac{F\pi}{2 \tau} & 0 & 0 \\
0 & 0 & 0 & 0 
\end{bmatrix} \text{ . } \label{EqLcdDaSimp}
\end{align}
Thus, the corresponding Lindbladian $\Lcal^{\text{DA}}_{\text{cd}}[\bullet]$ is
\begin{align}
\Lcal^{\text{DA}}_{\text{cd}}[\bullet] = \frac{1}{i\hbar} [H^{\text{DA}}_{\text{cd}},\bullet] \text{ , }
\end{align}
where we have the counter-diabatic Hamiltonian
\begin{align}
H^{\text{DA}}_{\text{cd}} = - \hbar \frac{F\pi}{4 \tau} \sigma_{z} \text{ . } \label{EqHcdDAOS}
\end{align}
Therefore, since the \textit{standard} TQD Lindbladian is given by $\Lmath^{\text{DA}}_{\text{Stqd}}(t) = \Lmath^{\text{DA}}(t)+\Lmath^{\text{DA}}_{\text{cd}}$, we conclude that
\begin{align}
\Lcal^{\text{DA}}_{\text{Stqd}}[\bullet] = \frac{1}{i\hbar} [H^{\text{DA}}_{\text{Stqd}}(t),\bullet] + \gamma(t) \left[ \sigma_{z} \bullet \sigma_{z} - \bullet \right] \text{ , } \label{EqLindTtqdDA}
\end{align}
with $H^{\text{DA}}_{\text{Stqd}}(t) = H^{\text{DA}}(t)+H^{\text{DA}}_{\text{cd}}$. Notice that, even though the counter-diabatic contribution for the Hamiltonian is time-independent, 
the total standard TQD Lindbladian depends on time through the Hamiltonian $H^{\text{DA}}(t)$. Moreover, the additional term $H^{\text{DA}}_{\text{cd}}$ to be introduced in 
Eq.~\eqref{EqLindDephDA}, which allows to implement the transitionless evolution in open system, does not depend on a reservoir engineering, but just on the fields that act on the system. 

\subsubsection{Deutsch generalized TQD}

Now, let us use the generalized approach of TQD in open systems to derive an alternative master equation for the Deutsch problem, which is 
time-independent but able to provide the same results as Eq.~\eqref{EqLindTtqdDA}. To this end, let us write the generalized TQD evolution operator as
\begin{align}
\Vcal_{\text{Gtqd}}^{\text{DA}}(t) = \sum_{\alpha = 0}^{N-1} e^{\int_{0}^{t} \Theta^{\text{DA}}_{\alpha}(\xi)d\xi} \dket{\Dcal_{\mu}^{\text{DA}}(t)}\dbra{\Ecal_{\mu}^{\text{DA}}(0)} \text{ , } 
\end{align}
where $\Theta^{\text{DA}}_{\alpha}(t)$ are the phases to be adjusted. However, as shown in Eq.~\eqref{EqIniStaDA}, the initial state of the system is written as a superposition of 
$\dket{\Dcal_{0}^{\text{DA}}(0)}$ and $\dket{\Dcal_{1}^{\text{DA}}(0)}$, so that the evolved state depends on the adiabatic phases $\Lambda_{0}(t)$ and $\Lambda_{1}(t)$, but it does 
not depend on the $\Lambda_{2}(t)$ and $\Lambda_{3}(t)$. Thus, in order to derive the generalized TQD, the generalized phases $\Theta^{\text{DA}}_{0}(t)$ and $\Theta^{\text{DA}}_{1}(t)$ 
should reproduce the adiabatic phases $\Lambda_{0}(t)$ and $\Lambda_{1}(t)$, but $\Theta^{\text{DA}}_{2}(t)$ and $\Theta^{\text{DA}}_{3}(t)$ allow us to introduce free parameters in the evolution.
In conclusion, by considering $\Theta^{\text{DA}}_{0}(t) = \Lambda^{\text{DA}}_{0}(t) = 0$ and $\Theta^{\text{DA}}_{1}(t)=\Lambda^{\text{DA}}_{1}(t)=-2\gamma(t)$, the generalized TQD Lindbladian 
is obtained from Eq.~\eqref{GenL1D} as 
\begin{align}
\Lmath^{\text{DA}(1)}_{\text{Gtqd}}(t) = 
\begin{bmatrix}
0 & 0 & 0 & 0 \\
0 & \eta_{1}(t) & \frac{\tau\tilde{\chi}+F\pi}{2\tau} & \frac{\omega \Theta_{23}^{-}(t)g_{\text{s}}(t)}{\Delta(t)} \\
0 & \frac{\tau\tilde{\chi}-F\pi}{2\tau} &
\eta_{2}(t) & \frac{\omega \Theta_{32}^{-}(t)g_{\text{c}}(t)}{\Delta(t)} \\
0 & -\frac{\omega \Theta_{32}^{-}(t)g_{\text{s}}(t)}{\Delta(t)} & -\frac{\omega \Theta_{32}^{-}(t)g_{\text{c}}(t)}{\Delta(t)} & \chi_{-}(t)
\end{bmatrix} \text{ , } \label{EqLGenSADaUnSimp}
\end{align}
where 
we have defined $\gamma(t) \equiv \gamma_{0}$ for a constant rate $\gamma_0$ (satisfying then $\dot{\gamma}(t)\!=\!0$), 
$\tilde{\chi}(t) = \sin(F\pi t) \left[ 2\gamma_{0} + \chi_{+}(t) \right]$, 
$\eta_{1}(t)\!=\!\Gamma_{+}(t) + g_{s}^2(t)\chi_{+}(t)$, $\eta_{2}(t)\!=\!\Gamma_{-}(t) + g_{c}^2(t)\chi_{+}(t)$, $\Theta_{32}^{-}(t)\!=\!\Theta_{3}(t)-\Theta_{2}(t)$, with
\begin{subequations}
	\label{EqChiDef}
	\begin{align}
	\chi_{\pm}(t) &= \left[\pm\Delta_{\pm}(t)\Theta_{3}(t)\mp\Delta_{\mp}(t)\Theta_{2}(t)\right]/(2\Delta(t)) , \\
	\Gamma_{\pm}(t) &= -\gamma_{0}(1 \pm g_{c}^{2}(t) \mp g_{s}^{2}(t)) \text{ . } 
	\end{align}
\end{subequations}
We can fix the free parameters at convenience so that we obtain a simple Lindblad superoperator. For instance, in order to get an anti-symmetric superoperator, 
we can adjust $\Theta_{3}(t)$. Indeed, from Eq.~\eqref{EqChiDef}, if we choose
\begin{align}
\Theta_{3}(t) = \frac{\Delta_{-}(t)\Theta_{2}(t) - 4 \gamma_{0} \Delta(t)}{\Delta_{+}(t)} , \label{EqTheta3DA}
\end{align}
we obtain an anti-symmetric Lindblad superoperator $\Lmath^{\text{DA}(2)}_{\text{Gtqd}}(t)$ given by
\begin{align}
\Lmath^{\text{DA}(2)}_{\text{Gtqd}}(t) = 
\begin{bmatrix}
0 & 0 & 0 & 0 \\
0 & -2\gamma_{0} & \frac{F\pi}{2\tau} & 
-\tilde{\Theta}_{2}(t)g_{\text{s}}(t) \\
0 & -\frac{F\pi}{2\tau} &
-2\gamma_{0} & -\tilde{\Theta}_{2}(t)g_{\text{c}}(t) \\
0 & \tilde{\Theta}_{2}(t)g_{\text{s}}(t) & \tilde{\Theta}_{2}(t)g_{\text{c}}(t) & 
-\frac{2\gamma_{0} \left(\Theta_{2}(t) - \Delta_{-}(t) \right)}{\Delta_{+}(t)}
\end{bmatrix} , \label{EqLGenSADaSymmetric}
\end{align}
where
\begin{eqnarray}
\tilde{\Theta}_{2}(t) = \frac{2\omega [2\gamma_{0}+\Theta_{2}(t)]}{\Delta_{+}(t)} \text{ . }
\end{eqnarray}
Concerning the parameter $\Theta_{2}(t)$, it is linked to reservoir and/or Hamiltonian engineering. For example, if we are not able to perform reservoir engineering, 
we take $\Theta_{2}(t) = \Lambda_{2}(t)$. In this case, we can show that the resulting generalized Lindblad superoperator $\Lmath^{\text{DA}(2)}_{\text{Gtqd}}(t)$ is 
given by $\Lmath^{\text{DA}(2)}_{\text{Gtqd}}(t) = \Lmath^{\text{DA}}_{\text{Ttqd}}(t) = \Lmath^{\text{DA}}(t)+\Lmath^{\text{DA}}_{\text{cd}}$, since 
$\Theta_{2}(t)$ and $\Theta_{3}(t)$ becomes exactly the phases which accompanies the adiabatic dynamics (see Eq.~\eqref{EqAdPhaDAOS}).

On the other hand, we can also consider others possibilities of choices for $\Theta_{2}(t)$ so that we can get alternative ways of driving the system. 
For example, from Eq.~\eqref{EqLGenSADaSymmetric}, we can identify some time-independent matrix elements of $\Lmath^{\text{DA}(2)}_{\text{Gtqd}}(t)$,  
with time-dependence on the parameter $\Theta_{2}(t)$ for other elements. In particular, it is worth highlighting that the right and left bases for $\Lmath^{\text{DA}}(t)$ satisfy 
Eq.~(\ref{EqTheo3}). Then, as a consequence of Theorem~\ref{TheoTimeIndL}, we can get a time-independent master equation. 
Particularly, it is obtained if we choose $\Theta_{2}(t) = -2\gamma_{0}$, so that the Lindbladian in Eq.~\eqref{EqLGenSADaSymmetric} becomes 
\begin{align}
\Lmath^{\text{DA}}_{\text{ti}} = 
\begin{bmatrix}
0 & 0 & 0 & 0 \\
0 & -2\gamma_{0} & \frac{F\pi}{2\tau} & 0 \\
0 & -\frac{F\pi}{2\tau} &
-2\gamma_{0} & 0 \\
0 & 0 & 0 & -2\gamma_{0}
\end{bmatrix} \text{ . } \label{EqLGenSADaTI}
\end{align}
For the associated master equation, we obtain
\begin{align}
\dot{\rho}(t) = \frac{1}{i\hbar} [H^{\text{DA}}_{\text{cd}},\rho(t)] + \Rcal^{\text{DA}}_{x}[\rho(t)] + \Rcal^{\text{DA}}_{y}[\rho(t)] + \Rcal^{\text{DA}}_{z}[\rho(t)] \text{ , } \label{EqLindGtqdDA}
\end{align}
with $H^{\text{DA}}_{\text{cd}}$ given by Eq.~\eqref{EqHcdDAOS} and $\Rcal^{\text{DA}}_{k}[\bullet] = \frac{\gamma_{0}}{2}(\sigma_{k}\bullet \sigma_{k} - \bullet)$. 
In order to verify whether Eqs.~\eqref{EqLindTtqdDA} and~\eqref{EqLindGtqdDA} allow us to get a shortcut to adiabatic Deutsch algorithm in open systems, we compute the fidelity~\cite{Nielsen:Book}
\begin{eqnarray}
\Fcal_{\text{os}}(\omega \tau) = \tr{\sqrt{\sqrt{\rho(\tau)}\rho^{\text{tar}}(\gamma_{0}\tau)\sqrt{\rho(\tau)}}} \text{ , } \label{EqFidelOS}
\end{eqnarray}
where $\rho(\tau)$ is a solution of Eq.~\eqref{EqLindTtqdDA} and~\eqref{EqLindGtqdDA} at $t=\tau$ and $\rho^{\text{tar}}$ is the target state. 
In our case, the target state is the adiabatic solution at the instant $t=\tau$ obtained from Eq.~\eqref{EqDAAdSolOS} as
\begin{align}
\rho^{\text{DA}}(\gamma_{0}\tau) &= \frac{1}{2} \left[ \1 + e^{-2\gamma_{0}\tau}\cos \left( \frac{\pi F}{2}\right) \sigma_{x} - e^{-2\gamma_{0}\tau}\sin \left( \frac{\pi F}{2} \right)\sigma_{y} \right]  \text{ . }\label{EqDAAdSolOSEnd}
\end{align}

\begin{figure}[t!]
	\centering
	\includegraphics[scale=0.27]{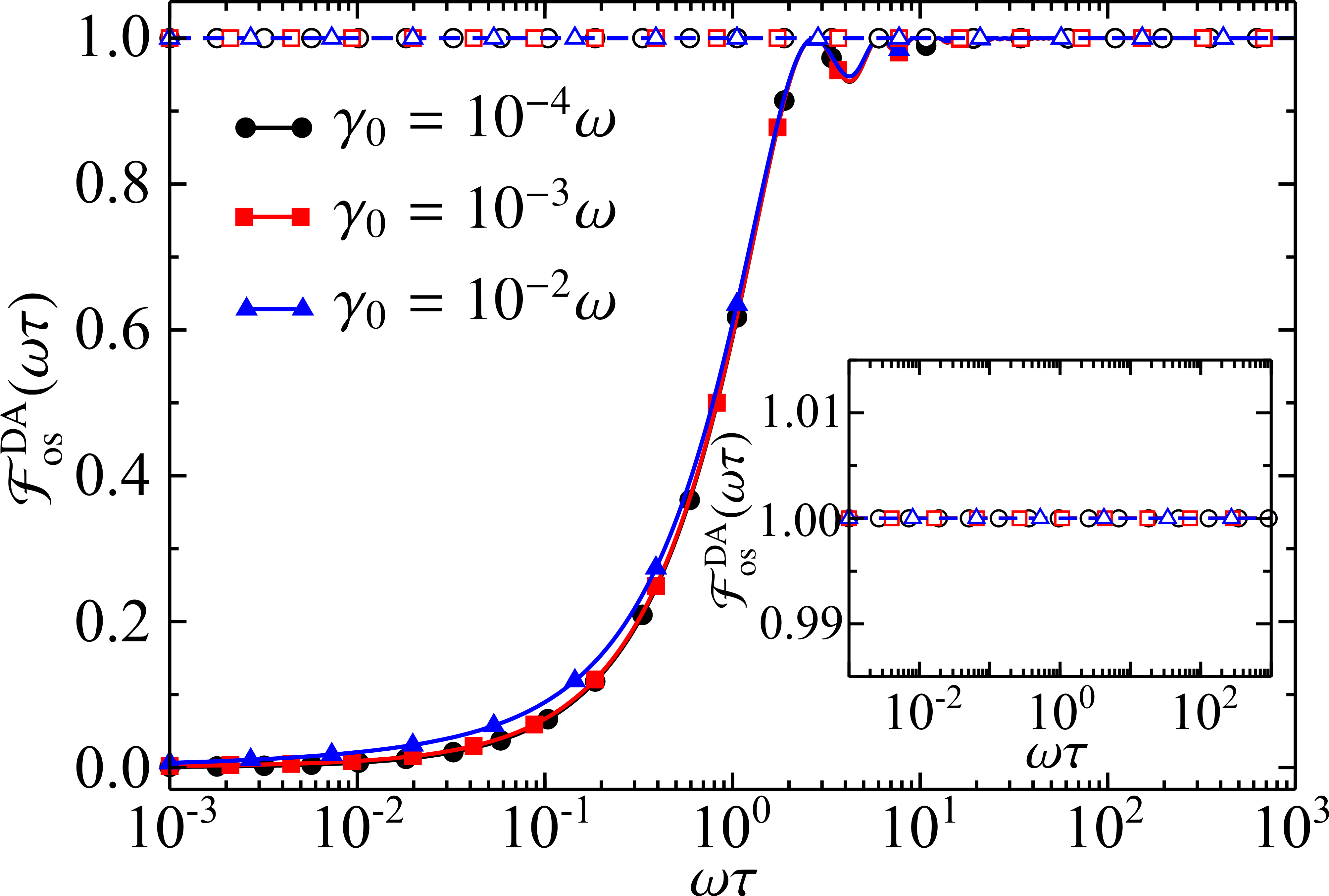}
	\caption{Fidelity $\Fcal^{\textrm{DA}}_{\text{os}}(\omega \tau)$ to achieve the open system adiabatic solution for the Deutsch algorithm for different values of $\gamma_{0}$. 
		Solid lines with filled symbols represent $\Fcal^{\textrm{DA}}_{\text{os}}(\omega \tau)$ when the system is driven by the master equation in Eq.~\eqref{EqLindDephDA}, 
		while dashed lines with empty symbols describe $\Fcal^{\textrm{DA}}_{\text{os}}(\omega \tau)$ when the system is driven by the generalized TQD master equation in Eq.~\eqref{EqLindGtqdDA}. 
		Inset: $\Fcal_{\text{os}}(\omega \tau)$ for the standard TQD evolution in Eq.~\eqref{EqLindTtqdDA}. Here we consider the case where the function is balanced.}
	\label{FigDATQD}
\end{figure}

\begin{figure}[t!]
	\centering
	\includegraphics[width=\linewidth]{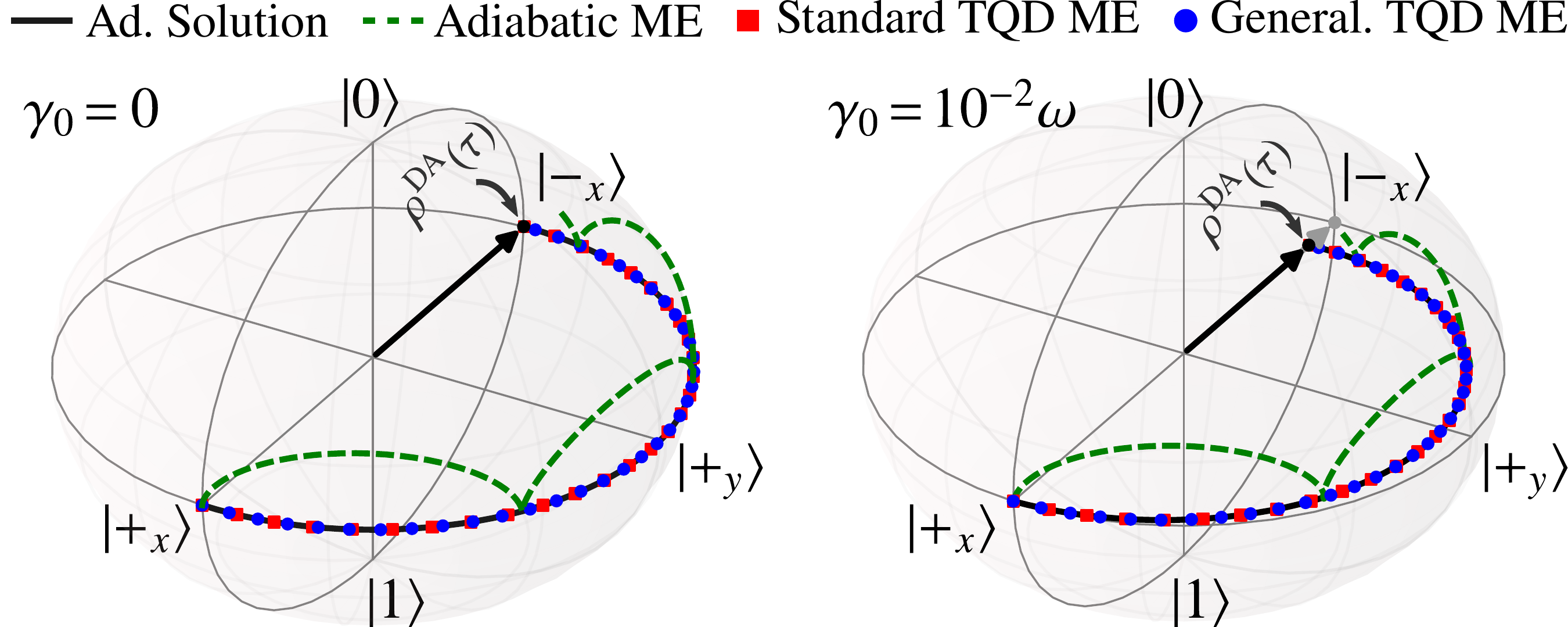}
	\caption{Trajectories in the Bloch sphere for several quantum evolutions: adiabatic solution given by Eq.~\eqref{EqDAAdSolOS} (Ad. Solution), exact solution of the master equation in Eq.~\eqref{EqLindDephDA} (Adiabatic ME), standard TQD approach in Eq.~\eqref{EqLindTtqdDA} (Standard TQD ME), and generalized TQD method in Eq.~\eqref{EqLindGtqdDA} (General. TQD ME). The dynamics is considered for closed (left Bloch sphere) and open 
	(right Bloch sphere) cases, where we consider the total evolution time such that $\tau\omega\!=\!10$.}
	\label{FigDASphere}
\end{figure}

In Fig.~\ref{FigDATQD} we present the fidelity as a function of $\omega \tau$, since we set $\gamma_{0}$ as a multiple of $\omega$. Both standard and generalized TQD protocols 
achieve a shortcut to adiabaticity in open system. It is important to highlight that a high fidelity here does not necessarily represent the optimal fidelity of the adiabatic Deutsch algorithm, 
because the state in Eq.~\eqref{EqDAAdSolOSEnd} is not an exact solution of the problem due to decoherence. This statement can be better understood from Fig.~\ref{FigDASphere}, 
where we show the trajectory of several distinct evolutions in the Bloch sphere. Regardless whether the system dynamics is unitary or not, as we drive the system in a time interval shorter 
than that required by the adiabatic conditions, the dynamics is far from the adiabatic solution, while both standard and generalized TQDs allow us to mimic the adiabatic behavior during all 
the time interval $t\!\in\![0,\tau]$. Notice also that, when the system is affected by decoherence, the state purity decreases, leading to a loss of fidelity to get the perfect output state.

\subsection{Landau-Zener model under bit-phase-flip}

As a second example of application of the generalized TQD approach, let us consider the Landau-Zener model, whose Hamiltonian is given by $
H_{\text{LZ}}(t) = (\hbar\omega_{0}/2) \sigma_{z} + (\hbar\Delta(t)/2) \sigma_{x}$. Here we are considering a time-independent detuning frequency $\omega_{0}$ and 
a time-dependent field $\Delta(t)$. 

\subsubsection{Landau-Zener adiabatic dynamics}

Let us assume that the system evolves under bit-phase flip decohering effect, whose the Lindblad equation is
\begin{align}
\dot{\rho}(t) = - \frac{i}{\hbar} [H_{\text{LZ}}(t),\rho(t)] + \gamma(t) \left[ \sigma_{y} \rho(t) \sigma_{y} - \rho(t) \right] \text{ , } \label{EqLindBitPPLZ}
\end{align}
where $\gamma(t)$ is the time-dependent bit phase flip decohering rate. From Eq.~(\ref{EqLindBitPPLZ}), we can write the corresponding superoperator $\Lmath^{\text{LZ}}(t)$ in the Pauli basis 
$\sigma_{i}=\{\1,\sigma_{x},\sigma_{y},\sigma_{z}\}$, yielding
\begin{align}
\Lmath^{\text{LZ}}(t) = \begin{bmatrix}
0 & 0 & 0 & 0 \\ 
0 & -2 \gamma(t) & -\omega_{0} & 0 \\ 
0 & \omega_{0} & 0 & -\omega_{0}\tan\theta(t) \\ 
0 & 0 & \omega_{0}\tan\theta(t) & -2 \gamma(t)
\end{bmatrix} \text{ , } 
\end{align}
where $\theta(t) = \arctan\left[\Delta(t)/\omega_{0}\right]$. The right eigenvectors are given by
\begin{subequations}
	\label{EqLZRightEigenVec}
	\begin{align}
	\dket{\Dcal^{\text{LZ}}_{0}(t)} &= \begin{bmatrix} \text{ } 1 & 0 & 0 & 0\text{ } \end{bmatrix}^{\text{t}} \text{ , } \\
	\dket{\Dcal^{\text{LZ}}_{1}(t)} &= \begin{bmatrix} \text{ } 0 & \sin \theta(t) & 0 & \cos \theta(t) \text{ } \end{bmatrix}^{\text{t}} \text{ , }
	\\
	\dket{\Dcal^{\text{LZ}}_{2}(t)} &= \begin{bmatrix} \text{ } 0 & -\cos \theta(t) & \frac{\gamma(t)\cos \theta(t) - \kappa(t)}{\omega_{0}} & \sin \theta(t) \text{ } \end{bmatrix}^{\text{t}} \text{ , }
	\\
	\dket{\Dcal^{\text{LZ}}_{3}(t)} &= \begin{bmatrix} \text{ } 0 & -\cos \theta(t) & \frac{\gamma(t)\cos \theta(t) + \kappa(t)}{\omega_{0}} & \sin \theta(t) \text{ } \end{bmatrix}^{\text{t}} \text{ , }
	\end{align}
\end{subequations}
while for the left eigenvectors, we have 
\begin{subequations}
	\label{EqLZLeftEigenVec}
	\begin{align}
	\dbra{\Ecal^{\text{LZ}}_{0}(t)} &= \begin{bmatrix} \text{ } 1 & 0 & 0 & 0\text{ } \end{bmatrix} \text{ , } \\
	\dbra{\Ecal^{\text{LZ}}_{1}(t)} &= \begin{bmatrix} \text{ } 0 & \sin \theta(t) & 0 & \cos \theta(t) \text{ } \end{bmatrix} \text{ , }
	\\
	\dbra{\Ecal^{\text{LZ}}_{2}(t)} &= \frac{1}{2} \begin{bmatrix} \text{ } 0 & -\cos \theta(t) \tilde{\kappa}_{+} & -\frac{\omega_{0}}{\kappa(t)} & \sin \theta(t) \tilde{\kappa}_{+} \text{ } \end{bmatrix} \text{ , }
	\\
	\dbra{\Ecal^{\text{LZ}}_{3}(t)} &= \frac{1}{2} \begin{bmatrix} \text{ } 0 & -\cos \theta(t) \tilde{\kappa}_{-} & \frac{\omega_{0}}{\kappa(t)} & \sin \theta(t) \tilde{\kappa}_{-} \text{ } \end{bmatrix} \text{ , }
	\end{align}
\end{subequations}
where we have defined $\tilde{\kappa}_{\pm} =  1 \pm \cos \theta(t) \gamma(t)/\kappa(t)$, $\kappa^2 (t) = \gamma^2(t)\cos^2 \theta(t)-\omega_0^2$, with eigenvalues $\lambda_{0}(t) = 0$, $\lambda_{1}(t) = -2 \gamma(t) $, $\lambda_{n}(t) = -\gamma(t) -(-1)^{n}\kappa(t)$, where $n=\{2,3\}$.

Now, by considering the case $\Delta(0)=0$ and by preparing the system in the ground state of $H(0)$, the initial state is given by $\rho^{\text{LZ}}(0) = \ket{1}\bra{1} = (1/2)(\1 - \sigma_{z})$. In the superoperator formalism we write
\begin{align}
\dket{\rho^{\text{LZ}}(0)} = \begin{bmatrix} \text{ } 1 & 0 & 0 & -1\text{ } \end{bmatrix}^{\text{t}} = \dket{\Dcal^{\text{LZ}}_{0}(0)} - \dket{\Dcal^{\text{LZ}}_{1}(0)} \text{ , }
\end{align}
where we already used that $\theta(0) = 0$ (since $\Delta(0)=0$) to write $\dket{\rho(0)}$ in terms of the eigenvectors of $\Lmath^{\text{LZ}}(0)$. If the system undergoes the adiabatic dynamics, its evolved state is given by
\begin{align}
\dket{\rho^{\text{LZ}}_{\text{ad}}(t)} = \dket{\Dcal^{\text{LZ}}_{0}(t)} - e^{\int_{t_{0}}^{t} \Lambda_{1}(\xi) d\xi}\dket{\Dcal^{\text{LZ}}_{1}(0)} \text{ , }
\end{align}
where $\Lambda_{1}(t)=\lambda_{1}(t)-\dinterpro{\Ecal_{1}(t)}{\dot{\Dcal}_{1}(t)}$. By using that $\dinterpro{\Ecal_{1}(t)}{\dot{\Dcal}_{1}(t)} = 0$, we then obtain $\Lambda_{1}(t)=-2\gamma(t)$. Thus, we write
\begin{align}
\dket{\rho^{\text{LZ}}_{\text{ad}}(t)} = \dket{\Dcal^{\text{LZ}}_{0}(t)} - e^{-\int_{t_{0}}^{t} 2 \gamma(\xi)d\xi}\dket{\Dcal^{\text{LZ}}_{1}(t)} \text{ . }
\end{align}
The explicit vector form of the above state is
\begin{align}
\dket{\rho^{\text{LZ}}_{\text{ad}}(t)} = \begin{bmatrix} \text{ } 1 & - e^{-\int_{t_{0}}^{t} 2 \gamma(\xi) d\xi}\sin \theta(t) & 0 & - e^{-\int_{t_{0}}^{t} 2 \gamma(\xi)d\xi} \cos \theta(t) \text{ } \end{bmatrix}^{\text{t}}
.
\end{align}
By turning back to the density matrix for the above dynamics, we then get
\begin{align}
\rho^{\text{LZ}}_{\text{ad}}(t) = \frac{1}{2} \left[\1 - e^{-\int_{t_{0}}^{t} 2 \gamma(\xi) d\xi}\sin \theta(t) \sigma_{x} - e^{-\int_{t_{0}}^{t} 2 \gamma(\xi)d\xi} \cos \theta(t)\sigma_{z} \right]
.
\end{align}

\subsubsection{Landau-Zener standard TQD}

From Eqs.~\eqref{EqLZRightEigenVec} and~\eqref{EqLZLeftEigenVec} we find the adiabatic phases associated with each eigenvector, which are given by
\begin{align}
\Lambda^{\text{LZ}}_{0} (t) &= 0 \text{ , } \quad \Lambda^{\text{LZ}}_{1} (t) = -2\gamma (t) \text{ , } \\  \Lambda^{\text{LZ}}_{2} (t) &= -\gamma (t) - \sec \theta(t) \kappa(t) -\Gcal^{\text{LZ}}_{2}(t) \text{ , } \\ \Lambda^{\text{LZ}}_{3} (t) &= -\gamma (t) + \sec \theta(t) \kappa(t) - \Gcal^{\text{LZ}}_{3}(t)\text{ , }
\end{align}
where $\Gcal^{\text{LZ}}_{\mu}(t)=\dinterpro{\Ecal_{\mu}^{\text{LZ}}(t)}{\dot{\Dcal}_{\mu}^{\text{LZ}}(t)}$ are the generalized Berry's phases
	\begin{subequations}
		\begin{align}
		\Gcal^{\text{LZ}}_{2}(t) &= \frac{\varpi(t)\left[ \kappa(t) - \cos \theta(t) \gamma (t) \right]}{2\kappa^2(t)\sec\theta(t)} , \label{gbp1} \\
		\Gcal^{\text{LZ}}_{3}(t) &= -\frac{\varpi(t)\left[ \kappa(t) + \cos \theta(t) \gamma (t) \right] }{2\kappa^2(t)\sec\theta(t)} ,  \label{gbp2} 
		\end{align}
	\end{subequations}
with $\varpi(t) = \gamma (t) \dot{\theta}(t) \tan \theta(t)  - \dot{\gamma} (t)$.
From Eqs.~(\ref{gbp1}) and (\ref{gbp2}), it is possible to see that, when $\varpi(t) = 0$, we obtain for the Landau-Zener model the generalized parallel transport condition $\Gcal^{\text{LZ}}_{\mu} (t)\!=\!0$, for all $\mu$. 
Notice that this is reached for 
\begin{align}
\gamma (t) = \tilde{\gamma} (t) = \gamma_{0}\sec \theta(t) , \label{EqGammaGenPTC}
\end{align}
for some constant $\gamma_{0}$. 
From the standard TQD theory, we can then write the counter-diabatic Lindbladian as
\begin{align}
\Lmath^{\text{LZ}}_{\text{cd}}(t) = \sum_{\mu=0}^{3} \left[\dket{\dot{\Dcal}_{\mu}^{\text{LZ}}(t)} \dbra{\Ecal_{\mu}^{\text{LZ}}(t)} -\Gcal^{\text{LZ}}_{\mu}(t) \dket{\Dcal_{\mu}^{\text{LZ}}(t)}\dbra{\Ecal_{\mu}^{\text{LZ}}(t)}\right] \text{ . }
\end{align}
By using Eqs.~\eqref{EqLZRightEigenVec} and~\eqref{EqLZLeftEigenVec} we find 
\begin{align}
\Lmath^{\text{LZ}}_{\text{cd}}(t) = \begin{bmatrix}
0 & 0 & 0 & 0 \\
0 & 0 & \frac{\tilde{\varpi}(t)}{2\kappa^2(t)} & \dot{\theta}(t) \\
0 & \frac{-3 \tilde{\varpi}(t)}{2\kappa^2(t)} & 0 & \frac{3\tilde{\varpi}(t) \sin \theta(t)}{2\cos \theta(t)\kappa^2(t)} \\
0 & -\dot{\theta}(t) & -\frac{\tilde{\varpi}(t) \sin \theta(t)}{2\cos \theta(t)\kappa^2(t)} & 0
\end{bmatrix} ,
\end{align}
where $\tilde{\varpi}\!=\!\omega_{0}\varpi(t) \cos^2 \theta(t)$. Similarly as discussed for the Deutsch problem, we can make the Lindblad superoperator $\Lmath^{\text{LZ}}_{\text{cd}}(t)$ anti-symmetric by imposing 
$\varpi(t) = 0$, which corresponds to the fulfillment of the generalized parallel transport condition, as provided by Eq.~\eqref{EqGammaGenPTC}. 
Hence, by using this condition, we get
\begin{align}
\Lmath^{\text{LZ}}_{\text{cd}}(t) = \begin{bmatrix}
0 & 0 & 0 & 0 \\
0 & 0 & 0 & \dot{\theta}(t) \\
0 & 0 & 0 & 0 \\
0 & -\dot{\theta}(t) & 0 & 0
\end{bmatrix}  \text{ . }
\end{align}
The Lindbladian contribution $\Lcal^{\text{LZ}}_{\text{cd}}[\bullet]$ associated with the counter-diabatic superoperator $\Lmath^{\text{LZ}}_{\text{cd}}(t)$ is then
\begin{align}
\Lcal^{\text{LZ}}_{\text{cd}}[\bullet] = \frac{1}{i\hbar} [H^{\text{LZ}}_{\text{cd}}(t),\bullet] \text{ , }
\end{align}
with the counter-diabatic Hamiltonian given by 
\begin{align}
H^{\text{LZ}}_{\text{cd}}(t) = \frac{\hbar \dot{\theta}(t)}{2} \sigma_{y} \text{ . } \label{EqHcdLZOS}
\end{align}
Therefore, the transitionless master equation is
\begin{align}
\dot{\rho}(t) = - \frac{i}{\hbar} [H^{\text{LZ}}_{\text{Stqd}}(t),\rho(t)] + \gamma (t) \left[ \sigma_{y} \rho(t) \sigma_{y} - \rho(t) \right] \text{ , } \label{EqLindTtqdLZ}
\end{align}
where $H^{\text{LZ}}_{\text{Stqd}}(t) = H^{\text{LZ}}(t) + H^{\text{LZ}}_{\text{cd}}(t)$. This correction term is exactly the counter-diabatic term for the Landau-Zener Hamiltonian found in Refs.~\cite{Santos:18-b,Hu:18}.  
In particular, it allows for the implementation of the adiabatic dynamics given by
\begin{align}
\tilde{\rho}^{\text{LZ}}_{\text{ad}}(t) = \frac{1}{2} \left[\1 - e^{-2\int_{0}^{t} \tilde{\gamma}(\xi) d\xi}\sin \theta(t) \sigma_{x} - e^{-2\int_{0}^{t} \tilde{\gamma}(\xi) d\xi} \cos \theta(t)\sigma_{z} \right]
. \label{EqRhoLZPart}
\end{align}

\subsubsection{Landau-Zener generalized TQD}

\begin{figure}
	\centering
	\includegraphics[scale=0.27]{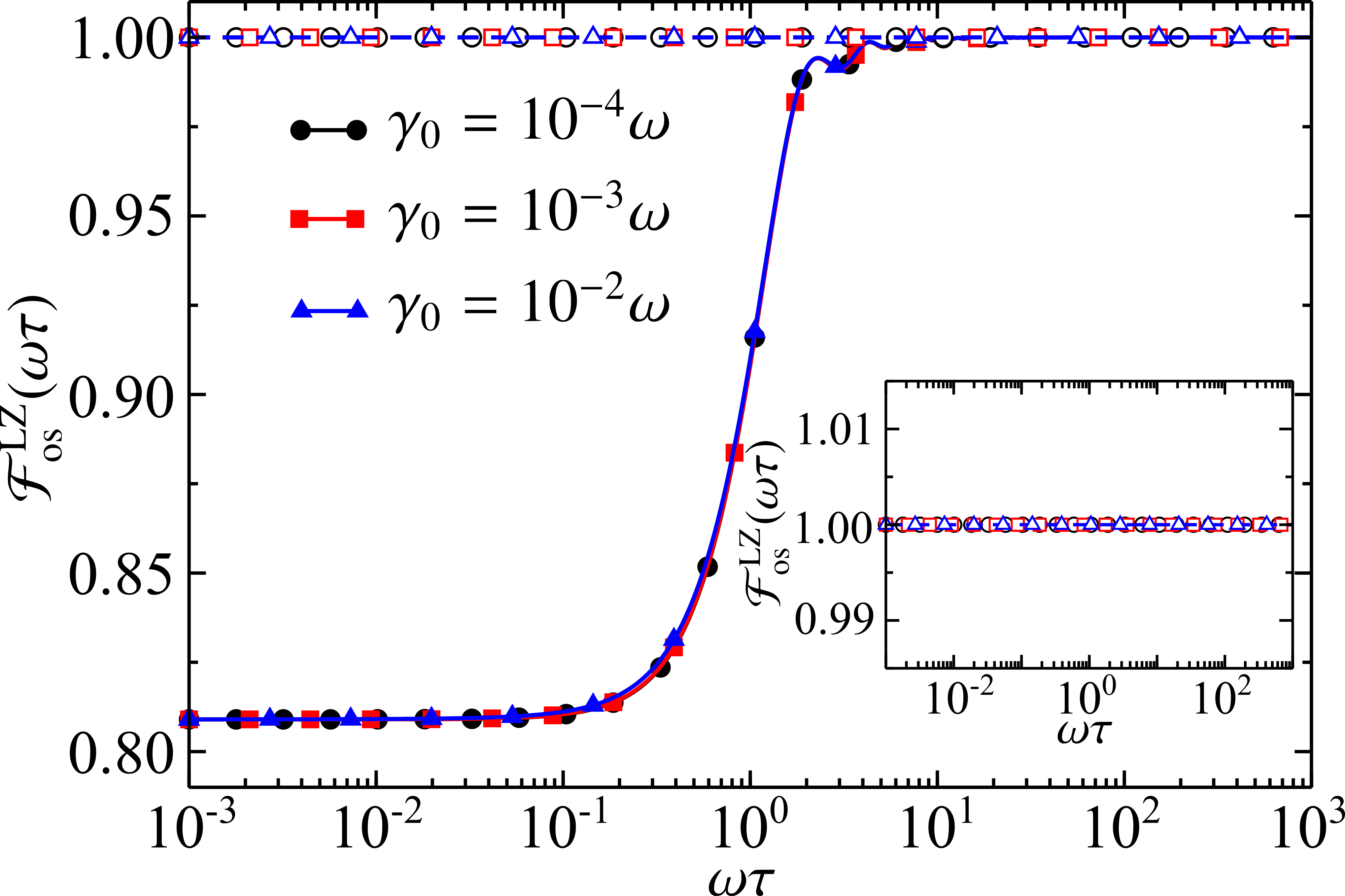}
	\caption{Fidelity $\Fcal^{\textrm{LZ}}_{\text{os}}(\omega \tau)$ to achieve the open system adiabatic target state of the Landau-Zener dynamics under bit-phase-flip for different values of $\gamma_{0}$. 
	Solid lines with filled symbols represent $\Fcal^{\textrm{LZ}}_{\text{os}}(\omega \tau)$ when the system is driven by the master equation in Eq.~\eqref{EqLindBitPPLZ}, while dashed lines with open symbols describe 
	$\Fcal^{\textrm{LZ}}_{\text{os}}(\omega \tau)$ when the system is driven by the generalized TQD master equation in Eq.~\eqref{EqLindGtqdLZ}. 
	Inset: $\Fcal_{\text{os}}(\omega \tau)$ for the standard TQD evolution in Eq.~\eqref{EqLindTtqdLZ}.}
	\label{FigLZTQD}
\end{figure}

\begin{figure}[t!]
	\centering
	\includegraphics[width=\linewidth]{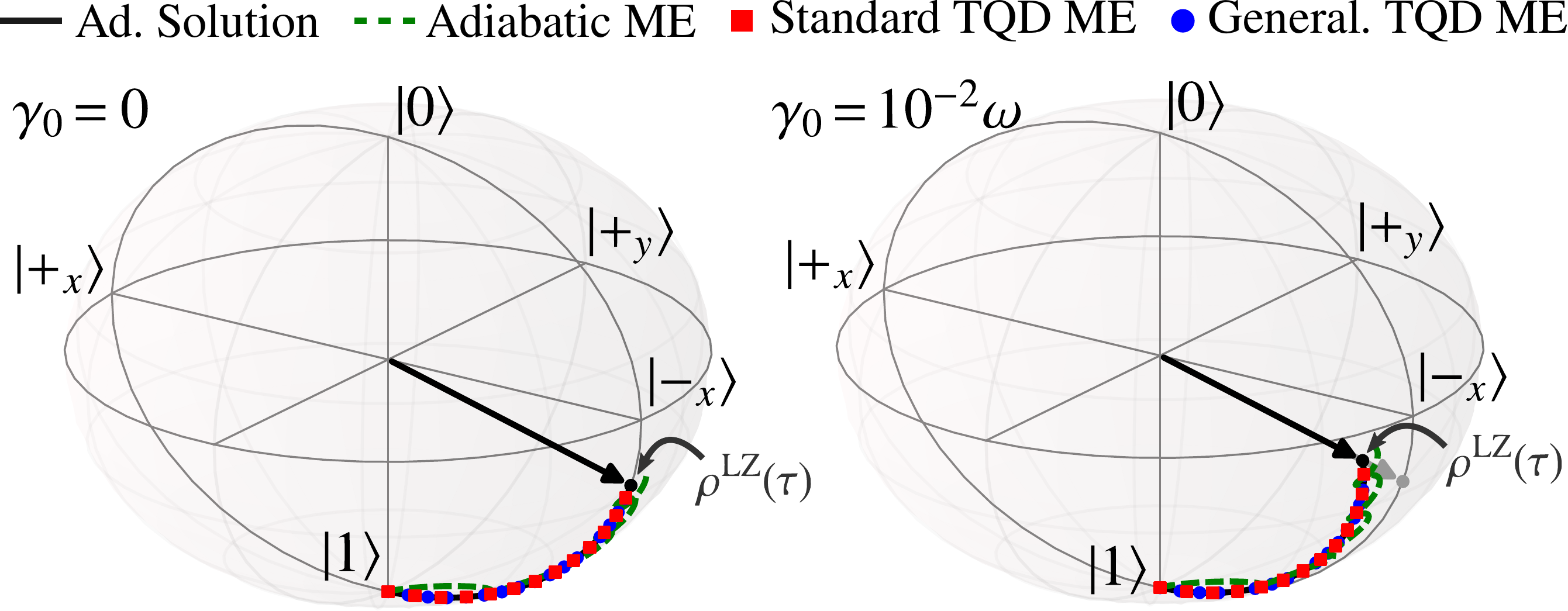}
	\caption{Trajectories in the Bloch sphere for several quantum evolutions: adiabatic solution given by Eq.~\eqref{EqRhoLZPart} (Ad. Solution), exact solution of the master equation in Eq.~\eqref{EqLindBitPPLZ} (Adiabatic ME), standard TQD approach in Eq.~\eqref{EqLindTtqdLZ} (Standard TQD ME), and generalized TQD method in Eq.~\eqref{EqLindGtqdLZ} (General. TQD ME). The dynamics is considered for closed (left Bloch sphere) and open 
	(right Bloch sphere) cases, where we consider the total evolution time such that $\tau\omega\!=\!10$.}
	\label{FigLZSphere}
\end{figure}

The generalized TQD for the Landau-Zener model is obtained by implementing the Lindblad superoperator
\begin{align}
\Lmath_{\text{Gtqd}}^{\text{LZ}}(t) = \sum_{\alpha = 0}^{3} \Theta^{\text{LZ}}_{\alpha}(t) \dket{\Dcal^{\text{LZ}}_{\alpha}(t)}\dbra{\Ecal^{\text{LZ}}_{\alpha}(t)} + \dket{\dot{\Dcal}^{\text{LZ}}_{\alpha}(t)} \dbra{\Ecal^{\text{LZ}}_{\alpha}(t)} \text{ . } \label{EqLindGenTQDLZ}
\end{align}
By employing the same procedure as performed for the Deutsch algorithm, we find that the generalized Lindblad superoperator can be written as
\begin{align}
\Lmath^{\text{LZ}}_{\text{Gtqd}}(t) = \begin{bmatrix}
0 & 0 & 0 & 0 
\\
0 & -2\gamma_{0}\sec \theta(t) & 0 & \tilde{\theta}
\\
0 & 0 & -2\gamma_{0}\sec \theta(t) & 0
\\
0 & -\tilde{\theta} & 0 & -2\gamma_{0}\sec \theta(t)
\end{bmatrix} \text{ , }
\end{align}
which is obtained by choosing $\Theta_{2}(t) = \Theta_{3}(t) = -2\gamma_{0} \sec \theta(t)$. The master equation then reads
\begin{align}
\dot{\rho}(t) = \frac{1}{i\hbar} [H^{\text{LZ}}_{\text{cd}},\rho(t)] + \Rcal^{\text{LZ}}_{x}[\rho(t)] + \Rcal^{\text{LZ}}_{y}[\rho(t)] + \Rcal^{\text{LZ}}_{z}[\rho(t)] \text{ , } \label{EqLindGtqdLZ}
\end{align}
where $H^{\text{LZ}}_{\text{cd}}$ is given by Eq.~\eqref{EqHcdLZOS} and 
\begin{align}
\Rcal^{\text{LZ}}_{k}[\bullet] = \frac{\gamma_{0}\sec \theta(t)}{2}(\sigma_{k}\bullet \sigma_{k} - \bullet) \text{ . }
\end{align}
This illustrates the fact that, in some situations, a time-independent Lindblad superoperator cannot be obtained, but the free parameters are still useful to provide time-independent fields for the Hamiltonian driving the system. 
As an example, let us consider $\theta(t) = \theta_{0} t/\tau$, so that Eq.~\eqref{EqRhoLZPart} provides, at $t = \tau$, the density operator
\begin{align}
\tilde{\rho}_{\text{ad}}^{\text{LZ}}(\tau) = \frac{1}{2} \left[\1 - e^{-2\gamma_{0}\tau \vartheta(\theta_{0})}\sin \theta_{0} \sigma_{x} - e^{-2\gamma_{0}\tau \vartheta(\theta_{0})} \cos \theta_{0}\sigma_{z} \right]
\text{ , }
\end{align}
where $\vartheta(\theta_{0}) = -\log\left( \cos \theta_{0}\right)/\theta_{0}$, for $0 \leq \theta_{0} < \pi/2$. In order to achieve this state, we plot the fidelity $\Fcal^{\text{LZ}}_{\text{os}}(\tau \omega)$ for the adiabatic 
and TQD dynamics in Fig~\ref{FigLZTQD}, where $\Fcal^{\text{LZ}}_{\text{os}}(\tau \omega)$ is computed from Eq.~\eqref{EqFidelOS}, with the target state given by $\tilde{\rho}^{\text{LZ}}_{\text{ad}}(\tau)$. As discussed in the previous section, the use of the TQD evolution to achieve high fidelity in open systems does not necessarily imply in high fidelity for the adiabatic trajectory in closed systems. 
This result is supported by Fig.~\ref{FigLZSphere}, where we illustrate several distinct trajectories in Bloch sphere.

\section{Conclusions}

We introduced a generalized approach for TQD in open systems, which provides a phase freedom in the Lindblad superoperator. Such phase freedom allows for an optimization of the 
evolution as we drive the system through a transitionless path. In particular, we discuss how to recover the standard TQD approach by suitably choosing the arbitrary phases of the generalized 
Lindblad superoperator. Moreover, we have provided a theorem providing sufficient conditions for a time-independent generalized TQD Lindblad superoperator. As a direct consequence, 
we can mimic the adiabatic dynamics in open systems using time-independent control for both Hamiltonian and reservoir engineering. 
Our results have been applied to two different situations. The first application concerns quantum computation through the adiabatic Deutsch algorithm under dephasing. For such evolution, we have determined 
the generalized phases so that we get a time-independent TQD master equation. As a second example, we have considered a two-level system driven by the Landau-Zener Hamiltonian, 
with interest in atomic/molecular physics. We have taken a nonunitary evolution under the bit-phase flip channel. Such example illustrates a time-dependent master equation, but with constant driving fields in the Hamiltonian. 
In both situations, we show how reservoir engineering can be used to 
obtain an open system parallel transport condition, which provides a geometric interpretation of shortcuts to adiabaticity in open systems. 

The methods presented here open perspectives for a number of applications in open system inverse engineering, with potential impact even to universal Lindblad-like master equations~\cite{Alipour:20}. 
For example, we can devise inverse engineering methods to design system-reservoir interactions able to track the ground state of a desired time-dependent Hamiltonian (induced adiabaticity). 
This is useful as an environment based quantum control technique, which resembles the approaches proposed in Refs.~\cite{Jing:16,Adolfo:20}.  

Extending the efficiency of quantum control to open systems is a forefront issue, which is still more challenging for the case of many-body real systems embedded in an external environment. Our approach offers 
finite-time flexible control as long as the system can be described by a tractable local master equation in the adiabatic regime and the required TQD interaction can be reduced to sufficiently local couplings. 
This is a point to be further explored in order to increase the size of quantum systems. 
Moreover, shortcuts to adiabaticity in open systems are also potentially fruitful to investigate quantum thermodynamics far from equilibrium. We have previously shown that the adiabatic behavior of open systems 
is compatible with the entropy variation at equilibrium~\cite{Santos:20c}. We can now envisage the application of our results to study entropy production in irreversible dynamics via counter-diabatic methods. 
Shortcuts to quantum thermalization and out of equilibrium processes via inverse engineering are expected to be addressed in future research.

\begin{acknowledgments}
A.C.S. acknowledges the financial support through the research grant from the São Paulo Research Foundation (FAPESP) (2019/22685-1). 
M.S.S. is supported by Conselho Nacional de Desenvolvimento Científico e Tecnológico (CNPq) (307854/2020-5).
This research is also supported in part by the Coordena\c{c}\~ao de Aperfei\c{c}oamento de Pessoal de N\'{\i}vel Superior - Brasil (CAPES) (Finance Code 001) and by the Brazilian
National Institute for Science and Technology of Quantum Information (INCT-IQ).
\end{acknowledgments}

\appendix

\section{Open systems in the superoperator formalism}\label{SecSuperOpForm}

In the theory of open systems, the dynamics of a quantum system $S$, with a Hilbert space $\Hcal_{\text{S}}$, coupled to an environment $A$, with a Hilbert space $\Hcal_{\text{A}}$, is generically described by a master equation that takes into account the system-environment interaction. In particular, we consider here a system dynamics described by the time-local master equation~\cite{Alicki:Book07,Petruccione:Book}
\begin{equation}
\dot{\rho}(t) = \Lcal_{t}[\rho(t)] \text{ , } \label{EqEqLind}
\end{equation}
where $\Lcal_{t}[\bullet]$ is the generator of the dynamics and the subscript ``$t$'' makes explicit the possibility of time-dependency of the parameters associated with the environment. One way to deal with the above equation is to define an extended space where the generator $\Lcal_{t}[\bullet]$ becomes a $(D^2 \times D^2)$-dimensional superoperator $\Lmath(t)$ and the density operator becomes a super-vector $\dket{\rho(t)}$. 
To see how it can be done, we define a set of $D^2 -1$ operators $\Ocal\!=\! \{\sigma_{n}\}\!\in\!\Hcal_{\text{S}}$ ($n\ge1$), so that $\trs{\sigma_{n}}\!=\!0$ and $\trs{\sigma_{n}\sigma_{m}^{\dagger}}\!=\!D\delta_{nm}$. The identity is introduced  as a $D$-dimensional operator $\sigma_{0}\!=\!\1$ so that we will be able to ensure $\trs{\rho(t)}\!=\!1$. In this form, we can expand $\rho(t)$ as
\begin{equation}
\rho = \frac{1}{D} \left[ \1 + \sum_{n=1}^{D^2-1} \varrho_{n} \sigma_{n} \right]\text{ , } \label{EqEqRhoCoherence}
\end{equation}
with $\varrho_{n}\!=\!\trs{\rho \sigma^{\dagger}_{n} }$. So, by using this expanded form of the density operator in Eq.~(\ref{EqEqLind}), we find the system of differential equations
\begin{equation}
\dot{\varrho}_{k} (t) = \frac{1}{D} \sum_{n=0}^{D^2 -1} \varrho_{i}(t) \trs{\sigma_{k}^{\dagger} \Lcal [ \sigma_{i} ]} \text{ , } \label{Eqv1}
\end{equation}
where we assume that $\Lcal [\bullet]$ is a linear operator. Note that if we identify the coefficient $\trs{\sigma_{k}^{\dagger} \Lcal [ \sigma_{i} ]}$ in the above equation as an element at $k$-th row and $i$-th column of a $(D^2 \times D^2)$-dimensional matrix $\Lmath(t)$, we can write
\begin{equation}
\dket{\dot{\rho}(t)} = \Lmath(t) \dket{\rho(t)} \text{ , } \label{EqEqSuperLindEq}
\end{equation}
where $\dket{\rho(t)}$ is a $D^2$-dimensional vector with components $\varrho_{n}(t) = \trs{\rho(t)\sigma_{n}^{\dagger}}$, $n=0,1,\cdots D^2 -1$. This proves Eq.~\eqref{LindEq} of the main text.

An important remark is that, due to the non-Hermiticity of the operator $\Lcal[\bullet]$, the superoperator $\Lmath(t)$ may not be diagonalizable. Nevertheless, square matrices can always be written in the Jordan canonical form, where $\Lmath(t)$ displays a block-diagonal structure $\Lmath_{\text{J}}(t)$ composed by Jordan blocks $J_{n}(t)$. The number of Jordan blocks is the sum of the geometric multiplicities of all the
eigenvalues $\lambda_{\alpha}(t)$ of $\Lmath(t)$~\cite{Horn:Book}. Then, we first define

\begin{definition}[Jordan block]\label{DefJordanForm}
	Given a $K\times K$ matrix $L$, the Jordan block form of $L$ reads
	\begin{align}
	L_{J} = \begin{bmatrix}
	J_{k_1}[\lambda_{k_1}] & 0                      & 0        & \cdots & 0 \\
	0                      & J_{k_2}[\lambda_{k_2}] & 0        & \cdots & 0 \\
	\vdots & \ddots & \ddots & \ddots & \vdots \\
	0 & \cdots & 0 & \ddots & 0  \\
	0 & \cdots & \cdots & 0 & J_{k_{N}}[\lambda_{N}] 
	\end{bmatrix}_{K \times K} \text{ , } \label{EqJordanFormMatrixJ}
	\end{align}
	where each block $J_{k_1}[\lambda_{k_1}]$ is given by an upper triangular matrix of the form
	\begin{align}
	J_{k}[\lambda] = \begin{bmatrix}
	\lambda & 1   & 0        & \cdots & 0 \\
	0 &\lambda & 1 & \cdots & 0 \\
	\vdots & \ddots & \ddots & \ddots & \vdots \\
	0 & \cdots & 0 & \lambda & 1  \\
	0 & \cdots & \cdots & 0 & \lambda  
	\end{bmatrix}_{k\times k} \text{ , } \label{EqJordanFormMatrix}
	\end{align}
	with $\lambda$ denoting the eigenvalues of $L$. Alternatively, a $K\times K$ Jordan matrix $L_{J}$ can be denoted as
	\begin{align}
	L_{J} = J_{k_{1}}[\lambda_{1}] \oplus J_{k_2}[\lambda_2] \oplus \cdots \oplus J_{k_N}[\lambda_N] = \bigoplus_{\alpha=1}^{N} J_{k_{\alpha}}[\lambda_{\alpha}] ,
	\end{align}
	where $N\leq K$ is the number of Jordan blocks required to write $L_{J}$ in a Jordan block form, and $k_1+k_2+\cdots+k_N = K$.
\end{definition}

In some cases, the coefficients $\lambda_n$ may depend on other parameters (time, for example). Then, by assuming that the coefficients $\lambda_n$ depend on a complete set of parameters $\xi = \{\xi_{1},\cdots \xi_{M}\}$, we denote the Jordan matrix as
\begin{align}
L_{J}(\xi) = J_{k_1}[\lambda_1(\xi)] \oplus J_{k_2}[\lambda_2(\xi)] \oplus \cdots \oplus J_{k_N}[\lambda_N(\xi)]  \text{ . }
\end{align}

The notion of Jordan form is important here because, different from the generator of a unitary dynamics (the Hamiltonian), the generator $\Lmath(t)$ in an open system does not always admit a diagonal form. 
However, every square matrix $A$ can be diagonalized by blocks from the \textit{Jordan canonical form theorem}~\cite{Horn:Book}.

\begin{theorem}[Jordan canonical form]
	Let $A \in M_{K}$ be $K \times K$ square matrix in the set $M_{K}$. Then, there is a non-singular matrix $S \in M_{K}$, positive integers $k_{1}, \cdots , k_{N}$, with $k_{1} + k_{2} + \cdots k_{N} = K$ ($N\leq K$), and scalars $\lambda_1, \cdots , \lambda_N \in \Cmath$ so that
	\begin{align}
	A(\xi) = S(\xi) J_{A}(\xi) S^{-1}(\xi) \text{ , }
	\end{align}
	where $J_{A}(\xi) = J_{k_1}[\lambda_1(\xi)] \oplus J_{k_2}[\lambda_2(\xi)] \oplus \cdots \oplus J_{k_N}[\lambda_N(\xi)]$ is the Jordan matrix associated with $A(\xi)$.
\end{theorem}

Now, by using the above discussion to the superoperator $\Lmath(t)$, we can obtain its Jordan form through the matrix $S(t)$, which allows us to write
\begin{equation}
\Lmath_{\text{J}}(t) = S^{-1}(t) \Lmath(t) S(t) = \bigoplus_{\alpha=1}^{N} L_{N_\alpha}[\lambda_{\alpha}(t)] \text{ , } \label{EqEqLindJ}
\end{equation}
where $N$ is the sum of the geometric multiplicities of all the
eigenvalues $\lambda_{\alpha}(t)$ of $\Lmath(t)$ and each block $L_{N_\alpha}[\lambda_{\alpha}(t)]$ is a $(N_\alpha \times N_\alpha)$-dimensional matrix given as in Eq.~\eqref{EqJordanFormMatrix}. Since the Hilbert space of the system has dimension $D$, one finds $N_1 + N_2 + \cdots + N_N = D^2$. In addition, as an immediate consequence of the structure of $\Lmath_{\text{J}}(t)$, we see that $\Lmath(t)$ does not always admit the existence of a complete set of eigenvectors. Instead, we define \textit{right} $\dket{\Dcal_{\alpha}^{n_{\alpha}}(t)}$ and \textit{left} $\dbra{\Ecal_{\alpha}^{n_{\alpha}}(t)}$ quasi-eigenvectors of $\Lmath(t)$ associated with the eigenvalue $\lambda_{\alpha}(t)$, satisfying
\begin{subequations}
	\begin{align}
	\Lmath(t)\dket{\Dcal_{\alpha}^{n_{\alpha}}(t)} &= \dket{\Dcal_{\alpha}^{(n_{\alpha}-1)}(t)} + \lambda_{\alpha}(t)\dket{\Dcal_{\alpha}^{n_{\alpha}}(t)} \text{ , } \\
	\dbra{\Ecal_{\alpha}^{n_{\alpha}}(t)}\Lmath(t) &= \dbra{\Ecal_{\alpha}^{(n_{\alpha}+1)}(t)} + \dbra{\Ecal_{\alpha}^{n_{\alpha}}(t)}\lambda_{\alpha}(t) \text{ . }
	\end{align}
\end{subequations}

The sets $\{\dket{\Dcal_{\alpha}^{n_{\alpha}}(t)}\}$ and $\{\dbra{\Ecal_{\alpha}^{n_{\alpha}}(t)}\}$ constitute bases for the space associated with the operator $\Lmath(t)$, satisfying 
the normalization condition $\dinterpro{\Ecal_{\beta}^{m_{\beta}}(t)}{\Dcal_{\alpha}^{n_\alpha}(t)} = \delta_{\beta\alpha} \delta_{m_{\beta}n_{\alpha}}$ and the completeness relationship
\begin{equation}
\sum_{\alpha=1}^{N} \sum _{n_{\alpha} = 1}^{N_{\alpha}} \dket{\Dcal_{\alpha}^{n_{\alpha}}(t)}\dbra{\Ecal_{\alpha}^{n_{\alpha}}(t)} = \1_{D^2\times D^2} \text{ , }
\end{equation}
where $N$ is the number of Jordan blocks in Eq.~\eqref{EqEqLindJ} and $N_{\alpha}$ is the dimension of the $\alpha$-th Jordan block.

\section{Standard counter-diabatic Lindbladians} \label{ApVacantiLCD}

Let us derive the standard counter-diabatic Lindblad superoperator from the similarity transformation 
$\Lmath_{\text{J}}(t) = C^{-1}(t) \Lmath(t) C(t)$ induced by the superoperator $C(t)$, with 
$\Lmath_{\text{J}}(t)$ denoting the Jordan canonical form of $\Lmath(t)$. From Eq.~(\ref{simtrans}), we have
\begin{eqnarray}
	\left[\Lmath_{\text{J}}(t) + \dot{C}^{-1}(t)C(t)\right] \dket{\rho(t)}_{\text{J}} = \dket{\dot{\rho}(t)}_{\text{J}} \text{ . }
	\label{A1}
\end{eqnarray}
with $\dket{\rho(t)}_{\text{J}}\!=\!C^{-1}(t)\dket{\rho(t)}_{\text{J}}$. We then split the term $\dot{C}^{-1}(t)C(t)$ into two contributions 
$\dot{C}^{-1}(t)C(t) = \Lmath_{\text{J}}^{\prime}(t) + \Lmath_{\text{nd}}^{\prime}(t)$, so that we can rewrite Eq.~(\ref{A1}) as
\begin{eqnarray}
	\left[\Lmath_{\text{J}}(t) + \Lmath_{\text{J}}^{\prime}(t) + \Lmath_{\text{nd}}^{\prime}(t)\right] \dket{\rho(t)}_{\text{J}} = \dket{\dot{\rho}(t)}_{\text{J}} \text{ , }
	\label{A2}
\end{eqnarray}
where
\begin{align}
\Lmath_{\text{J}}^{\prime}(t) &= \sum_{\mu=0}^{N-1}\sum_{n_{\mu}=1}^{N_{\mu}}\sum_{\ell_{\mu}=1}^{N_{\mu}} C_{\mu\mu}^{n_{\mu}\ell_{\mu}}(t)
\dket{\sigma_{\mu}^{n_{\mu}}}\dbra{\sigma_{\mu}^{\ell_{\mu}}} \text{ , } \\
\Lmath_{\text{nd}}^{\prime}(t) &= \sum_{\nu\neq\mu}^{N-1}\sum_{n_{\nu}=1}^{N_{\nu}}\sum_{\mu=0}^{N-1}\sum_{\ell_{\mu}=1}^{N_{\mu}}
C_{\nu\mu}^{n_{\nu}\ell_{\mu}}(t)\dket{\sigma_{\nu}^{n_{\nu}}}\dbra{\sigma_{\mu}^{\ell_{\mu}}} \text{ , }
\end{align}
with coefficients $C_{\nu\mu}^{n_{\nu}\ell_{\mu}}(t) = \dbra{\sigma_{\nu}^{n_{\nu}}} \dot{C}^{-1}(t)C(t)\dket{\sigma_{\mu}^{\ell_{\mu}}}$. 
Eq.~(\ref{A2}) leads to the counter-diabatic Lindbladian as proposed by Vacanti \textit{et al.} in Ref.~\cite{Vacanti:14}, which reads
\begin{eqnarray}
\Lmath_{\text{cd}}(t) &= - C(t)\Lmath_{\text{nd}}^{\prime}(t)C^{-1}(t) \text{ , }
\end{eqnarray}
Henceforth, we aim at providing $\Lmath_{\text{cd}}(t)$ in terms of the right and left quasi-eigenbasis $\{\dket{\Dcal_{\beta}^{m_{\beta}}(t)}\}$  and $\{\dbra{\Ecal_{\beta}^{m_{\beta}}(t)}\}$. 
In this direction, we start by expressing $\Lmath_{\text{cd}}(t)$ as
\begin{align}
\Lmath_{\text{cd}}(t) &= - C(t)\Lmath_{\text{nd}}^{\prime}(t)C^{-1}(t) \nonumber \\
&\hspace{-0.9cm} = - C(t)\Lmath_{\text{nd}}^{\prime}(t)C^{-1}(t) + 
\left[ C(t)\Lmath_{\text{J}}^{\prime}(t)C^{-1}(t) - C(t)\Lmath_{\text{J}}^{\prime}(t)C^{-1}(t)\right] \nonumber \\
&\hspace{-0.9cm}= - C(t)\left[\Lmath_{\text{nd}}^{\prime}(t)+ \Lmath_{\text{J}}^{\prime}(t)\right]C^{-1}(t) + 
C(t)\Lmath_{\text{J}}^{\prime}(t)C^{-1}(t) \nonumber \\
&\hspace{-0.9cm}= - C(t)\left[\dot{C}^{-1}(t)C(t)\right]C^{-1}(t) + 
C(t)\Lmath_{\text{J}}^{\prime}(t)C^{-1}(t) .
\end{align}
Let us now use $\dot{C}^{-1}(t)C(t) = -C^{-1}(t)\dot{C}(t)$, yielding
\begin{eqnarray}
\Lmath_{\text{cd}}(t) &= \dot{C}(t)C^{-1}(t) + 
C(t)\Lmath_{\text{J}}^{\prime}(t)C^{-1}(t) \text{ . }
\label{apA}
\end{eqnarray}
Thus, we can compute each term in the right-hand-side of Eq.~(\ref{apA}) as
\begin{align}
\dot{C}(t)C^{-1}(t) &= \left[\sum_{\mu=0}^{N-1}\sum_{n_{\mu}=1}^{N_{\mu}} \dket{\dot{\Dcal}_{\mu}^{n_{\mu}}(t)} \dbra{\sigma_{\mu}^{n_{\mu}}}\right]
\left[\sum_{\nu=0}^{N-1}\sum_{k_{\nu}=1}^{N_{\nu}} \dket{\sigma_{\nu}^{k_{\nu}}} \dbra{\Ecal_{\nu}^{k_{\nu}}(t)}\right] 
\nonumber \\
&= \sum_{\mu=0}^{N-1}\sum_{n_{\mu}=1}^{N_{\mu}} \dket{\dot{\Dcal}_{\mu}^{n_{\mu}}(t)} \dbra{\Ecal_{\mu}^{n_{\mu}}(t)} \text{ , }
\label{apACC}
\end{align}
and
\begin{widetext}
\begin{align}
C(t)\Lmath_{\text{J}}^{\prime}(t)C^{-1}(t) &= \left[\sum_{\mu=0}^{N-1}\sum_{n_{\mu}=1}^{N_{\mu}} \dket{\Dcal_{\mu}^{n_{\mu}}(t)} \dbra{\sigma_{\mu}^{n_{\mu}}}\right]
\left[
\sum_{\eta=0}^{N-1}\sum_{j_{\eta}=1}^{N_{\eta}}\sum_{\ell_{\eta}=1}^{N_{\eta}} C_{\eta\eta}^{j_{\eta}\ell_{\eta}}(t)
\dket{\sigma_{\eta}^{j_{\eta}}}\dbra{\sigma_{\eta}^{\ell_{\eta}}}
\right]
\left[\sum_{\nu=0}^{N-1}\sum_{k_{\nu}=1}^{N_{\nu}} \dket{\sigma_{\nu}^{k_{\nu}}} \dbra{\Ecal_{\nu}^{k_{\nu}}(t)}\right]  \nonumber \\
&= \sum_{\mu=0}^{N-1}\sum_{n_{\mu}=1}^{N_{\mu}} \sum_{\eta=0}^{N-1}\sum_{j_{\eta}=1}^{N_{\eta}}\sum_{\ell_{\eta}=1}^{N_{\eta}}
\sum_{\nu=0}^{N-1}\sum_{k_{\nu}=1}^{N_{\nu}} 
\left[ \dket{\Dcal_{\mu}^{n_{\mu}}(t)} \dbra{\sigma_{\mu}^{n_{\mu}}}\right]
\left[ C_{\eta\eta}^{j_{\eta}\ell_{\eta}}(t)\dket{\sigma_{\eta}^{j_{\eta}}}\dbra{\sigma_{\eta}^{\ell_{\eta}}}\right]
\left[\dket{\sigma_{\nu}^{k_{\nu}}} \dbra{\Ecal_{\nu}^{k_{\nu}}(t)}\right]
\nonumber \\
&= \sum_{\mu=0}^{N-1}\sum_{n_{\mu}=1}^{N_{\mu}} \sum_{\eta=0}^{N-1}\sum_{j_{\eta}=1}^{N_{\eta}}\sum_{\ell_{\eta}=1}^{N_{\eta}}
\sum_{\nu=0}^{N-1}\sum_{k_{\nu}=1}^{N_{\nu}} 
\left[ C_{\eta\eta}^{j_{\eta}\ell_{\eta}}(t)\dket{\Dcal_{\mu}^{n_{\mu}}(t)} 
\underbrace{\dinterpro{\sigma_{\mu}^{n_{\mu}}}{\sigma_{\eta}^{j_{\eta}}} }_{\delta_{\mu\eta} \delta_{n_{\mu}j_{\eta}}}
\underbrace{\dinterpro{\sigma_{\eta}^{\ell_{\eta}}}{\sigma_{\nu}^{k_{\nu}}}}_{\delta_{\nu\eta} \delta_{\ell_{\eta}k_{\nu}}}
\dbra{\Ecal_{\nu}^{k_{\nu}}(t)}\right]
\nonumber \\
&= \sum_{\mu=0}^{N-1}\sum_{n_{\mu}=1}^{N_{\mu}} \sum_{k_{\mu}=1}^{N_{\mu}} 
\left[ C_{\mu\mu}^{n_{\mu}k_{\mu}}(t)\dket{\Dcal_{\mu}^{n_{\mu}}(t)} 
\dbra{\Ecal_{\mu}^{k_{\mu}}(t)}\right] = \sum_{\mu=0}^{N-1}\sum_{n_{\mu},k_{\mu}=1}^{N_{\mu}} 
\left[ \dbra{\sigma_{\mu}^{n_{\mu}}} \dot{C}^{-1}(t)C(t)\dket{\sigma_{\mu}^{k_{\mu}}}\dket{\Dcal_{\mu}^{n_{\mu}}(t)}\dbra{\Ecal_{\mu}^{k_{\mu}}(t)}\right] ,
\label{apA2} 
\end{align}
\end{widetext}
where we used $C_{\mu\mu}^{n_{\mu}k_{\mu}}(t) = \dbra{\sigma_{\mu}^{n_{\mu}}} \dot{C}^{-1}(t)C(t)\dket{\sigma_{\mu}^{k_{\mu}}}$ in the last equality. 
Now, we use the definition of $C(t)$ in Eq.~(\ref{Csupop}) to express $\dot{C}^{-1}(t)C(t)$ as 
\begin{equation}
\dot{C}^{-1}(t)C(t) = -\sum_{\eta=0}^{N-1}\sum_{\ell_{\eta}=1}^{N_{\eta}} \sum_{\nu=0}^{N-1}\sum_{j_{\nu}=1}^{N_{\nu}} \Gcal_{\eta\nu}^{\ell_{\eta}j_{\nu}}(t)
\dket{\sigma_{\eta}^{\ell_{\eta}}}\dbra{\sigma_{\nu}^{j_{\nu}}} \text{ , }
\label{apA3}
\end{equation}
where $\Gcal_{\eta\nu}^{\ell_{\eta}j_{\nu}}(t)\!=\!\dinterpro{\Ecal_{\eta}^{\ell_{\eta}}(t)}{\dot{\Dcal}_{\nu}^{j_{\nu}}(t)}$, so that
\begin{align}
\dbra{\sigma_{\mu}^{n_{\mu}}}\dot{C}^{-1}(t)C(t)\dket{\sigma_{\mu}^{k_{\mu}}} &= 
-\sum_{\eta,\nu=0}^{N-1}\sum_{\ell_{\eta,j_{\nu}}=1}^{N_{\eta},N_{\nu}}
\Gcal_{\eta\nu}^{\ell_{\eta}j_{\nu}}(t)
\delta_{\mu\eta} \delta_{n_{\mu}\ell_{\eta}}
\delta_{\nu\mu} \delta_{j_{\nu}k_{\mu}}
\nonumber \\ 
&= -\Gcal_{\mu\mu}^{n_{\mu}k_{\mu}}(t) 
. \label{APG}
\end{align}
Therefore, by inserting Eq.~(\ref{APG}) in Eq.~(\ref{apA2}), we get
\begin{align}
C(t)\Lmath_{\text{J}}^{\prime}(t)C^{-1}(t) &= -\sum_{\mu=0}^{N-1}\sum_{n_{\mu},k_{\mu}=1}^{N_{\mu}} 
\left[ \Gcal_{\mu\mu}^{n_{\mu}k_{\mu}}(t) \dket{\Dcal_{\mu}^{n_{\mu}}(t)}\dbra{\Ecal_{\mu}^{k_{\mu}}(t)}\right] \text{ . }
\label{apACLC}
\end{align}
So, from Eqs.~(\ref{apA}), (\ref{apACC}), and (\ref{apACLC}), we conclude that
\begin{align}
\Lmath_{\text{cd}}(t) = \sum_{\mu=0}^{N-1}\sum_{n_{\mu}=1}^{N_{\mu}} &\left[\dket{\dot{\Dcal}_{\mu}^{n_{\mu}}} \dbra{\Ecal_{\mu}^{n_{\mu}}} - \sum_{k_{\mu}=1}^{N_{\mu}}\Gcal_{\mu\mu}^{n_{\mu}k_{\mu}} \dket{\Dcal_{\mu}^{n_{\mu}}}\dbra{\Ecal_{\mu}^{k_{\mu}}} \right] \text{ . }
\end{align}
Therefore, $\Lmath_{\text{cd}}(t)$ exhibits a structure that is formally identical to the standard counter-diabatic Hamiltonian $H_{\text{cd}}(t)$ in Eq.~\eqref{HtqdCStheory}. Let us now show that $\Lmath_{\text{cd}}(t)$ mimics the adiabatic evolution in open systems. In the superoperator formalism, the master equation is given by
\begin{eqnarray}
\dket{\dot{\rho}(t)} &= \left[ \Lmath(t) + \Lmath_{\text{cd}}(t)\right]\dket{\rho(t)} \text{ . }
\end{eqnarray}
We now use the expansion $\dket{\rho(t)}\!=\!\sum_{\alpha,n_{\alpha}}r_{\alpha}^{n_{\alpha}}(t)\dket{\Dcal_{\alpha}^{n_{\alpha}}(t)}$ in terms of the right quasi-eigenbasis of $\Lmath(t)$. 
For each component $r_{\alpha}^{n_{\alpha}}(t)\dket{\Dcal_{\alpha}^{n_{\alpha}}(t)}$ we then get
\begin{align}
\frac{d}{dt}[ r_{\alpha}^{n_{\alpha}}(t)\dket{\Dcal_{\alpha}^{n_{\alpha}}(t)}] = r_{\alpha}^{n_{\alpha}}(t) \left[ \Lmath(t) + \Lmath_{\text{cd}}(t)\right] \dket{\Dcal_{\alpha}^{n_{\alpha}}(t)} .
\label{apAr}
\end{align}
By defining $\Lmath_{\text{Stqd}}(t)\!=\!\Lmath(t) + \Lmath_{\text{cd}}(t)$, we obtain
\begin{align}
\hspace{-0.3cm}\Lmath_{\text{Stqd}}(t) \dket{\Dcal_{\alpha}^{n_{\alpha}}(t)} &= \left[ \Lmath(t) + \Lmath_{\text{cd}}(t)\right] \dket{\Dcal_{\alpha}^{n_{\alpha}}(t)} \nonumber \\
&=\dket{\Dcal_{\alpha}^{(n_{\alpha}-1)}(t)} + \lambda_{\alpha}(t)\dket{\Dcal_{\alpha}^{n_{\alpha}}(t)} \nonumber \\
&+ \sum_{\mu=0}^{N-1}\sum_{n_{\mu}=1}^{N_{\mu}} \dket{\dot{\Dcal}_{\mu}^{n_{\mu}}(t)} \dinterpro{\Ecal_{\mu}^{n_{\mu}}(t)}{\Dcal_{\alpha}^{n_{\alpha}}(t)} \nonumber \\
&- \sum_{\mu=0}^{N-1}\sum_{n_{\mu},k_{\mu}=1}^{N_{\mu}} \Gcal_{\mu\mu}^{n_{\mu}k_{\mu}}(t)  \dket{\Dcal_{\mu}^{n_{\mu}}(t)} \dinterpro{\Ecal_{\mu}^{k_{\mu}}(t)}{\Dcal_{\alpha}^{n_{\alpha}}(t)} \nonumber \\
&\hspace{-1.6cm}= \dket{\Dcal_{\alpha}^{(n_{\alpha}-1)}} +  \lambda_{\alpha} \dket{\Dcal_{\alpha}^{n_{\alpha}}} + \dket{\dot{\Dcal}_{\alpha}^{n_{\alpha}}}  - \sum_{n_{\mu}=1}^{N_\mu} \Gcal_{\alpha\alpha}^{n_{\mu}n_{\alpha}} 
\dket{\Dcal_{\alpha}^{n_{\mu}}} ,
\end{align}
so that, after projecting Eq.~(\ref{apAr}) over $\dbra{\Ecal_{\beta}^{j_{\beta}}(t)}$, the dynamics for the coefficient $r_{\beta}^{j_{\beta}}(t)$ reads 
\begin{align}
\dot{r}_{\beta}^{j_{\beta}}(t) =  \lambda_{\beta}(t)  r_{\beta}^{j_{\beta}}(t) -  \sum_{n_{\alpha}=1}^{N_\alpha} \Gcal_{\beta\beta}^{j_\beta n_{\alpha}}(t) \, r_{\beta}^{n_{\alpha}}(t) + r_{\beta}^{(j_{\beta}+1)}(t)   \text{ . }
\label{apAdr}
\end{align}
Eq.~(\ref{apAdr}) implies in the adiabatic behavior, with the Jordan blocks decoupled from each other and the standard open-system adiabatic phase fixed~\cite{Sarandy:05-1,Santos:20c}. 
Therefore, the Lindbladian $\Lmath_{\text{Stqd}}(t)\!=\!\Lmath^{\prime}(t)\!=\!\Lmath(t) + \Lmath_{\text{cd}}(t)$ exactly mimics the adiabatic dynamics in open systems. 

\section{Inverse engineering in open system} \label{ApenIEOS}

Let us start from the time-local master equation provided by Eq.~\eqref{LindEq} and assume that the evolved density operator 
$\dket{\rho(t)}$ can be obtained by a non-unitary superoperator $ \Vcal \left( t,t_{0}\right)$ as 
$\dket{\rho(t)} = \Vcal \left( t,t_{0}\right) \dket{\rho(t_{0})}$. Then
\begin{eqnarray}
\dot{\Vcal} \left( t,t_{0}\right)\dket{\rho(t_{0})} = \Lmath(t)\Vcal \left( t,t_{0}\right) \dket{\rho(t_{0})} \text{ , }
\end{eqnarray}
which holds for any initial state $\dket{\rho(t_{0})}$. Therefore
\begin{eqnarray}
\dot{\Vcal} \left( t,t_{0}\right) = \Lmath(t)\Vcal \left( t,t_{0}\right) \text{ . } \label{PreLinver}
\end{eqnarray}
We assume an open system evolution driven by an invertible dynamical map. Then, we consider that a superoperator $\Vcal^{-1} \left( t,t_{0}\right)$ so that $\Vcal^{-1} \left( t,t_{0}\right)\Vcal \left( t,t_{0}\right) = \1$. 
By multiplying Eq.~\eqref{PreLinver} by $\Vcal^{-1} \left( t,t_{0}\right)$, we get
\begin{eqnarray}
\Lmath(t) = \dot{\Vcal} \left( t,t_{0}\right)\Vcal^{-1} \left( t,t_{0}\right) \text{ . } \label{ApLinver}
\end{eqnarray}
Eq.~(\ref{ApLinver}) generalizes the inverse engineering approach to the realm of open quantum systems.  

\section{Generalized TQD for 1D Jordan decomposition} \label{ApRecTTQDL}

Let $\Lmath(t)$ be a Lindbladian superoperator that admits one-dimensional Jordan-block decomposition. Then, the phase-free generalized TQD evolution operator reads 
\begin{eqnarray}
\Vcal_{\text{Gtqd}}^{1\text{D}}(t,t_{0}) = \sum_{\alpha = 0}^{N-1} e^{\int_{t_{0}}^{t} \Theta_{\alpha}(\xi)d\xi} \dket{\Dcal_{\alpha}(t)}\dbra{\Ecal_{\alpha}(t_{0})} \text{ , }
\end{eqnarray}
so that the associated Lindbladian is
\begin{align}
\Lmath_{\text{Gtqd}}^{1\text{D}}(t) = \sum_{\alpha = 0}^{N-1}\sum_{\beta = 0}^{N-1} &\frac{d}{dt}\left[e^{\int_{t_{0}}^{t} \Theta_{\alpha}(\xi)d\xi} \dket{\Dcal_{\alpha}(t)}\right]\dbra{\Ecal_{\beta}(t)}  \nonumber \\
&\times e^{-\int_{t_{0}}^{t} \Theta_{\beta}(\xi)d\xi}\dinterpro{\Ecal_{\alpha}(t_{0})}{\Dcal_{\beta}(t_{0})} \nonumber \\
= \sum_{\alpha = 0}^{N-1} \frac{d}{dt}&\left[e^{\int_{t_{0}}^{t} \Theta_{\alpha}(\xi)d\xi} \dket{\Dcal_{\alpha}(t)}\right]\dbra{\Ecal_{\alpha}(t)} e^{-\int_{t_{0}}^{t} \Theta_{\alpha}(\xi)d\xi} \nonumber \\
= \sum_{\alpha = 0}^{N-1} \Theta_{\alpha}&(t) e^{\int_{t_{0}}^{t} \Theta_{\alpha}(\xi)d\xi} \dket{\Dcal_{\alpha}(t)}\dbra{\Ecal_{\alpha}(t)} e^{-\int_{t_{0}}^{t} \Theta_{\alpha}(\xi)d\xi}  \nonumber \\
&+e^{\int_{t_{0}}^{t} \Theta_{\alpha}(\xi)d\xi} \dket{\dot{\Dcal}_{\alpha}(t)}\dbra{\Ecal_{\alpha}(t)} e^{-\int_{t_{0}}^{t} \Theta_{\alpha}(\xi)d\xi}\nonumber \\ = \sum_{\alpha = 0}^{N-1} \Theta_{\alpha}&(t) \dket{\Dcal_{\alpha}(t)}\dbra{\Ecal_{\alpha}(t)} + \dket{\dot{\Dcal}_{\alpha}(t)} \dbra{\Ecal_{\alpha}(t)} \text{ . }
\end{align}
The standard TQD Lindbladian $\Lmath_{\text{Stqd}}^{1\text{D}}(t)$ can be recovered by imposing that $\Theta_{\alpha}(t)$ is equal to the open-system adiabatic phase, 
{\it i.e.}, $\Theta_{\alpha}(t) = \lambda_{\alpha}(t) - \dinterpro{\Ecal_{\alpha}(t)}{\dot{\Dcal}_{\alpha}(t)}$. In that case, we have
\begin{align}
\Lmath_{\text{Gtqd}}^{1\text{D}}(t) &= \overbrace{\sum_{\alpha = 0}^{N-1}\lambda_{\alpha}(t) \dket{\Dcal_{\alpha}(t)}\dbra{\Ecal_{\alpha}(t)}}^{\Lmath(t)} \nonumber \\&+ \underbrace{\sum_{\alpha=0}^{N-1} \dket{\dot{\Dcal}_{\alpha}(t)} \dbra{\Ecal_{\alpha}(t)} -\dinterpro{\Ecal_{\alpha}(t)}{\dot{\Dcal}_{\alpha}(t)} \dket{\Dcal_{\alpha}(t)}\dbra{\Ecal_{\alpha}(t)}}_{\Lmath_{\text{cd}}^{1\text{D}}(t)}  \nonumber \\&= \Lmath(t) + \Lmath_{\text{cd}}^{\text{1D}}(t)  = \Lmath_{\text{Stqd}}^{1\text{D}}(t)\text{ . }
\end{align}

\section{Proof of the Theorem~\ref{TheoTimeIndL}} \label{ApTheoTimeIndL}

We now demonstrate Theorem~\ref{TheoTimeIndL}. To this end, we need to consider the time derivative of $\Lmath_{\text{Gtqd}}^{1\text{D}}(t)$, which yields
\begin{align}
\dot{\Lmath}_{\text{Gtqd}}^{1\text{D}}(t) = \frac{d}{dt}&\left[\sum_{\alpha = 0}^{N-1} \Theta_{\alpha}(t) \dket{\Dcal_{\alpha}(t)}\dbra{\Ecal_{\alpha}(t)} + \dket{\dot{\Dcal}_{\alpha}(t)} \dbra{\Ecal_{\alpha}(t)}\right] \nonumber \\
= \sum_{\alpha = 0}^{N-1} &\left[\dot{\Theta}_{\alpha}(t) \dket{\Dcal_{\alpha}(t)}\dbra{\Ecal_{\alpha}(t)} + \Theta_{\alpha}(t) \dket{\dot{\Dcal}_{\alpha}(t)}\dbra{\Ecal_{\alpha}(t)} \right. \nonumber \\
&\left. + \Theta_{\alpha}(t) \dket{\Dcal_{\alpha}(t)}\dbra{\dot{\Ecal}_{\alpha}(t)} \right]  \nonumber \\ +\sum_{\alpha = 0}^{N-1}& \left[\dket{\ddot{\Dcal}_{\alpha}(t)} \dbra{\Ecal_{\alpha}(t)} + \dket{\dot{\Dcal}_{\alpha}(t)} \dbra{\dot{\Ecal}_{\alpha}(t)}\right] \text{ . }
\end{align}
By computing the matrix elements $\dot{\Lmath}_{\text{Gtqd}}^{1\text{D}}(t)\vert_{\eta \beta}\!=\!\dbra{\Ecal_{\eta}(t)}\dot{\Lmath}_{\text{Gtqd}}^{1\text{D}}(t) \dket{\Dcal_{\beta}(t)}$, we obtain
\begin{align}
\dot{\Lmath}_{\text{Gtqd}}^{1\text{D}}(t)\vert_{\eta \beta} = 
\sum_{\alpha = 0}^{N-1} &\left[\dot{\Theta}_{\alpha}(t) \delta_{\eta\alpha}\delta_{\alpha\beta} + \Theta_{\alpha}(t) \dinterpro{\Ecal_{\eta}(t)}{\dot{\Dcal}_{\alpha}(t)} \delta_{\alpha\beta} + \right. \nonumber \\ & \left. \Theta_{\alpha}(t) \delta_{\eta\alpha}\dinterpro{\dot{\Ecal}_{\alpha}(t)}{\Dcal_{\beta}(t)} \right]  \nonumber \\
+ \sum_{\alpha = 0}^{N-1}& \left[\dinterpro{\Ecal_{\eta}(t)}{\dot{\Dcal}_{\alpha}(t)} \dinterpro{\dot{\Ecal}_{\alpha}(t)}{\Dcal_{\beta}(t)} + \right. \nonumber \\ &\left. \dinterpro{\Ecal_{\eta}(t)}{\ddot{\Dcal}_{\alpha}(t)} \delta_{\alpha\beta} \right] \nonumber \\
= 
\dot{\Theta}_{\eta}&(t) \delta_{\eta\beta} + [\Theta_{\beta}(t) - \Theta_{\eta}(t)] \dinterpro{\Ecal_{\eta}(t)}{\dot{\Dcal}_{\beta}(t)}   \nonumber \\ +\dinterpro{\Ecal_{\eta}&(t)}{\ddot{\Dcal}_{\beta}(t)} + 
\dinterpro{\dot{\Ecal}_{\eta}(t)}{\dot{\Dcal}_{\beta}(t)} \text{ , }
\end{align}
where we have used the identity $\dinterpro{\Ecal_{\eta}(t)}{\dot{\Dcal}_{\alpha}(t)} = - \dinterpro{\dot{\Ecal}_{\eta}(t)}{\Dcal_{\alpha}(t)}$ (due to the left-right eigenvector orthonormalization) 
and the completeness relation $\sum_{\alpha = 0}^{N-1} \dket{\Dcal_{\alpha}(t) }\dbra{\Ecal_{\alpha}(t)} = \1$. 
Thus, in order to get a time-independent Lindbladian superoperator, we need to find parameters $\Theta_{\beta}(t)$ so that $\dot{\Lmath}_{\text{Gtqd}}^{1\text{D}}(t) = 0$. This occurs if, and only if, we require
\begin{align}
\dot{\Lmath}_{\text{Gtqd}}^{1\text{D}}(t)\vert_{\eta \eta} &= \dot{\Theta}_{\eta}(t) + \frac{d}{dt}\left[\dinterpro{\Ecal_{\eta}(t)}{\dot{\Dcal}_{\eta}(t)}\right] = 0 \text{ , } \label{ApEqThetaOpt1}\\
\dot{\Lmath}_{\text{Gtqd}}^{1\text{D}}(t)\vert_{\eta,\beta\neq\eta} &= [\Theta_{\beta}(t) - \Theta_{\eta}(t)] \dinterpro{\Ecal_{\eta}(t)}{\dot{\Dcal}_{\beta}(t)} \nonumber \\ & 
+ \frac{d}{dt}\left[\dinterpro{\Ecal_{\eta}(t)}{\dot{\Dcal}_{\beta}(t)}\right] =  0 \text{ . } \label{ApEqThetaOpt2}
\end{align}
Let us assume we have 
\begin{align}
\frac{d}{dt}\left[\dinterpro{\Ecal_{\eta}(t)}{\dot{\Dcal}_{\beta}(t)}\right] = 0 \text{ , } \forall \eta, \beta \label{ConditionTheta}
\end{align}
Under this condition, it is possible to solve Eqs.~(\ref{ApEqThetaOpt1}) and~(\ref{ApEqThetaOpt2}) by simply imposing
\begin{eqnarray}
\Theta_{\beta} = \Theta_{\eta} = \text{constant} \text{ , }  \forall \eta, \beta
\end{eqnarray}
On the other hand, even if Eq.~\eqref{ConditionTheta} is not satisfied, we may find an analytical solution. 
By explicitly solving Eq.~\eqref{ApEqThetaOpt1} for $\Theta_{\eta}(t)$ and by denoting the solution as $\overline{\Theta}_{\eta}(t)$, we obtain
\begin{align}
\overline{\Theta}_{\eta}(t) - \overline{\Theta}_{\eta}(t_0) &= - \int_{t_{0}}^{t} \frac{d}{d\xi}\left[\dinterpro{\Ecal_{\eta}(\xi)}{\dot{\Dcal}_{\eta}(\xi)}\right]d\xi \nonumber \\ &= \dinterpro{\Ecal_{\eta}(t_0)}{\dot{\Dcal}_{\eta}(t_0)} - \dinterpro{\Ecal_{\eta}(t)}{\dot{\Dcal}_{\eta}(t)}\text{ , }
\end{align}
so that
\begin{eqnarray}
\overline{\Theta}_{\eta}(t) = - \dinterpro{\Ecal_{\eta}(t)}{\dot{\Dcal}_{\eta}(t)} \equiv \Gcal_{\eta}(t) \text{ . }
\label{ApDSol}
\end{eqnarray}
However, we still have to further impose Eq.~\eqref{ApEqThetaOpt2} in order to guarantee that $\overline{\Theta}_{\eta}(t)$ provides a time-independent Lindbladian. 
By using Eq.~(\ref{ApDSol}) in Eq.~\eqref{ApEqThetaOpt2}, we get
\begin{align}
\dot{\Lmath}_{\text{Gtqd}}^{1\text{D}}(t)\vert_{\eta,\beta\neq\eta} &= [\overline{\Theta}_{\beta}(t) - \overline{\Theta}_{\eta}(t)] \dinterpro{\Ecal_{\eta}(t)}{\dot{\Dcal}_{\beta}(t)}\nonumber \\
& + \frac{d}{dt}\left[\dinterpro{\Ecal_{\eta}(t)}{\dot{\Dcal}_{\beta}(t)}\right] \nonumber \\
&= [\Gcal_{\beta}(t) - \Gcal_{\eta}(t) ] \dinterpro{\Ecal_{\eta}(t)}{\dot{\Dcal}_{\beta}(t)} \nonumber \\
&+ \frac{d}{dt}\left[\dinterpro{\Ecal_{\eta}(t)}{\dot{\Dcal}_{\beta}(t)}\right]  = 0 \text{ . }
\end{align}
Therefore, we have a differential equation to $\dinterpro{\Ecal_{\eta}(t)}{\dot{\Dcal}_{\beta}(t)}$, which can be rewritten as 
\begin{equation}
\frac{\frac{d}{dt}\left[\dinterpro{\Ecal_{\eta}(t)}{\dot{\Dcal}_{\beta}(t)}\right]}{\dinterpro{\Ecal_{\eta}(t)}{\dot{\Dcal}_{\beta}(t)}} = \Gcal_{\eta}(t) -\Gcal_{\beta}(t)  \text{ . }
\label{eqD}
\end{equation}
Eq.~(\ref{eqD}) then yields
\begin{equation}
\frac{d}{dt} \ln \left[ \dinterpro{\Ecal_{\eta}(t)}{\dot{\Dcal}_{\beta}(t)}\right] =\Gcal_{\eta}(t) - \Gcal_{\beta}(t)  \text{ , }
\end{equation}
where, by integrating, we find
\begin{align}
\ln  \left[ \frac{\dinterpro{\Ecal_{\eta}(t)}{\dot{\Dcal}_{\beta}(t)}}{\dinterpro{\Ecal_{\eta}(t_{0})}{\dot{\Dcal}_{\beta}(t_{0})}} \right] =  \int_{t_{0}}^{t} 
\left[ \Gcal_{\eta}(\xi)  - \Gcal_{\beta}(\xi)  \right] d\xi \text{ . }
\end{align}
Therefore, we conclude that the condition to be satisfied is given by
\begin{align}
\dinterpro{\Ecal_{\eta}(t)}{\dot{\Dcal}_{\beta}(t)} = \dinterpro{\Ecal_{\eta}(t_{0})}{\dot{\Dcal}_{\beta}(t_{0})} \, e^{\int_{t_{0}}^{t} \left[\Gcal_{\eta}(\xi)  - \Gcal_{\beta}(\xi) \right] d\xi }\, ,  \,\,\, \forall \eta, \beta \text{. }
\end{align}
We can now analyze a particular case. As discussed in Ref.~\cite{Hu:19-a}, from the set of right-eigenvectors of the Lindbladian, it is always possible to get a time-independent eigenvector $\dket{\Dcal_{0}(t)} = \dket{\Dcal_{0}}$ 
associated with an eigenvalue $\lambda_{0} = 0$. Let us consider the matrix elements of the Lindbladian for this constant eigenvector. From Eqs.~(\ref{ApEqThetaOpt1}) and~(\ref{ApEqThetaOpt2}), we write
\begin{align}
\dbra{\Ecal_{0}(t)}\dot{\Lmath}_{\text{Gtqd}}^{1\text{D}}(t) \dket{\Dcal_{0}(t)} &= \dot{\Theta}_{0}(t) + \frac{d}{dt}\left[\dinterpro{\Ecal_{0}(t)}{\dot{\Dcal}_{0}(t)}\right] = 0 \text{ , } \\
\dbra{\Ecal_{0}(t)}\dot{\Lmath}_{\text{Gtqd}}^{1\text{D}}(t) \dket{\Dcal_{\beta\neq 0}(t)} &= [\Theta_{\beta}(t) - \Theta_{0}(t)] \dinterpro{\Ecal_{0}(t)}{\dot{\Dcal}_{\beta}(t)} \nonumber \\ &+ \frac{d}{dt}\left[\dinterpro{\Ecal_{0}(t)}{\dot{\Dcal}_{\beta}(t)}\right] = 0 \text{ . }
\end{align}
Therefore, since $\dket{\Dcal_{0}(t)} = \dket{\Dcal_{0}}$, we can always choose a corresponding time-independent left-eigenvector $\dbra{\Ecal_{0}(t)} = \dbra{\Ecal_{0}}$, so that $\dinterpro{\Ecal_{0}(t)}{\dot{\Dcal}_{\beta}(t)}=0$, $\forall \beta$ (due to the orthonormalization condition). Hence
\begin{align}
\dbra{\Ecal_{0}(t)}\dot{\Lmath}_{\text{Gtqd}}^{1\text{D}}(t) \dket{\Dcal_{0}(t)} &= \dot{\Theta}_{0}(t) = 0 \text{ , } \\
\dbra{\Ecal_{0}(t)}\dot{\Lmath}_{\text{Gtqd}}^{1\text{D}}(t) \dket{\Dcal_{\beta\neq 0}(t)} &= 0 \text{ . }
\end{align}
This shows that $\Theta_{0}(t)$ can be chosen to be constant independently of the remaining parameters $\Theta_{\beta\neq 0}(t)$, {\it i.e.}, the definition of the phase specifically associated with the vanishing eigenvalue $\lambda_0 = 0$ is 
decoupled from the other sectors $\beta \ne 0$.   

\section{Generalized TQD Lindblad superoperator for multi-dimensional Jordan blocks} \label{ApGeneralizedLCD}

Let us obtain the generalized TQD Lindbladian for multi-dimensional Jordan blocks, which is provided by Eq.~\eqref{LG_tqd}. First, we write the evolution superoperator, which reads 
\begin{eqnarray}
\Vcal_{\text{Gtqd}}(t,t_{0}) = \sum_{\alpha = 0}^{N-1} \sum _{n_{\alpha} = 1}^{N_{\alpha}} \sum _{m_{\alpha} = 1}^{N_{\alpha}} q_{\alpha}^{n_{\alpha}m_{\alpha}}\dket{\Dcal_{\alpha}^{n_{\alpha}}(t)}\dbra{\Ecal_{\alpha}^{m_{\alpha}}(t_{0})} \text{ , }
\end{eqnarray}
where the time-dependence of the coefficients $q$'s is omitted. Then, we can write the inverse evolution superoperator $\Vcal_{\text{TQD}-1}^{\text{Gen}}(t,t_{0})$ as
\begin{eqnarray}
\Vcal_{\text{Gtqd}}^{-1}(t,t_{0}) = \sum_{\alpha = 0}^{N-1} \sum _{n_{\alpha} = 1}^{N_{\alpha}} \sum _{m_{\alpha} = 1}^{N_{\alpha}} \tilde{q}_{\alpha}^{n_{\alpha}m_{\alpha}}\dket{\Dcal_{\alpha}^{n_{\alpha}}(t_{0})}\dbra{\Ecal_{\alpha}^{m_{\alpha}}(t)} \text{ , }
\end{eqnarray}
where the coefficients $\tilde{q}_{\alpha}^{n_{\alpha}m_{\alpha}}$ are such that $\Vcal_{\text{Gtqd}}(t,t_{0})\Vcal_{\text{Gtqd}}^{-1}(t,t_{0})=\1$. In order to determine the conditions to be obeyed by 
$q_{\alpha}^{n_{\alpha}m_{\alpha}}$ and $\tilde{q}_{\alpha}^{n_{\alpha}m_{\alpha}}$, let us explicitly consider the expression
\begin{widetext}
\begin{align}
\Vcal_{\text{Gtqd}}(t,t_{0})\Vcal_{\text{Gtqd}}^{-1}(t,t_{0}) &= \sum_{\alpha = 0}^{N-1} \sum _{n_{\alpha} = 1}^{N_{\alpha}} \sum _{m_{\alpha} = 1}^{N_{\alpha}}
\sum_{\beta = 0}^{N-1} \sum _{k_{\beta} = 1}^{N_{\beta}} \sum _{\ell_{\beta} = 1}^{N_{\beta}}
q_{\alpha}^{n_{\alpha}m_{\alpha}} \tilde{q}_{\beta}^{k_{\beta}\ell_{\beta}}
\dinterpro{\Ecal_{\alpha}^{m_{\alpha}}(t_{0})}{\Dcal_{\beta}^{k_{\beta}}(t_{0})} \dket{\Dcal_{\alpha}^{n_{\alpha}}(t)}\dbra{\Ecal_{\beta}^{\ell_{\beta}}(t)} \nonumber \\
&= \sum_{\alpha = 0}^{N-1} \sum _{n_{\alpha} = 1}^{N_{\alpha}} \sum _{m_{\alpha} = 1}^{N_{\alpha}}
\sum _{\ell_{\alpha} = 1}^{N_{\alpha}}
q_{\alpha}^{n_{\alpha}m_{\alpha}} \tilde{q}_{\alpha}^{m_{\alpha}\ell_{\alpha}} \dket{\Dcal_{\alpha}^{n_{\alpha}}(t)}\dbra{\Ecal_{\alpha}^{\ell_{\alpha}}(t)} \text{ , }
\end{align}
so that we write
\begin{align}
\dbra{\Ecal_{\eta}^{l_{\eta}}(t)}\Vcal_{\text{Gtqd}}(t,t_{0})\Vcal_{\text{Gtqd}}^{-1}(t,t_{0})\dket{\Dcal_{\kappa}^{i_{\kappa}}(t)} &= \sum_{\alpha = 0}^{N-1} \sum _{n_{\alpha} = 1}^{N_{\alpha}} \sum _{m_{\alpha} = 1}^{N_{\alpha}}
\sum _{\ell_{\alpha} = 1}^{N_{\alpha}}
q_{\alpha}^{n_{\alpha}m_{\alpha}} \tilde{q}_{\alpha}^{m_{\alpha}\ell_{\alpha}} \dinterpro{\Ecal_{\eta}^{l_{\eta}}(t)}{\Dcal_{\alpha}^{n_{\alpha}}(t)}\dinterpro{\Ecal_{\alpha}^{\ell_{\alpha}}(t)}{\Dcal_{\kappa}^{i_{\kappa}}(t)} = \delta_{\eta\kappa}\sum _{m_{\kappa} = 1}^{N_{\kappa}}
q_{\kappa}^{l_{\kappa}m_{\kappa}} \tilde{q}_{\kappa}^{m_{\kappa}i_{\kappa}} \text{ . }
\end{align}
\end{widetext}
Thus, we can get $\Vcal_{\text{Gtqd}}(t,t_{0})\Vcal_{\text{Gtqd}}^{-1}(t,t_{0})=\1$ by imposing
\begin{eqnarray}
\sum _{m_{\kappa} = 1}^{N_{\kappa}}
q_{\kappa}^{l_{\kappa}m_{\kappa}} \tilde{q}_{\kappa}^{m_{\kappa}i_{\kappa}} = \delta_{l_{\kappa}i_{\kappa}} \label{ApEqqDelta}\text{ . }
\end{eqnarray}
Now, we can derive the generalized TQD Lindbladian $\Lmath_{\text{Gtqd}}(t,t_{0}) $, which implements the dynamics governed by the evolution superoperator $\Vcal_{\text{Gtqd}}(t,t_{0})$. From Eq.~(\ref{EqLGTQDfromU}), we obtain
\begin{eqnarray}
\Lmath_{\text{Gtqd}}(t,t_{0})  &=& \dot{\Vcal}_{\text{Gtqd}} \left( t,t_{0}\right) \Vcal_{\text{Gtqd}}^{-1} \left( t,t_{0}\right) \nonumber \\
&=& \sum_{\alpha = 0}^{N-1} \sum _{n_{\alpha} = 1}^{N_{\alpha}}  \sum _{k_{\alpha} = 1}^{N_{\alpha}} \sum _{l_{\alpha} = 1}^{N_{\alpha}} \left(
\dot{q}_{\alpha}^{n_{\alpha}k_{\alpha}}\tilde{q}_{\alpha}^{k_{\alpha}l_{\alpha}}\dket{\Dcal_{\alpha}^{n_{\alpha}}(t)}\dbra{\Ecal_{\alpha}^{l_{\alpha}}(t)} \right. \nonumber \\
&&\left. \hspace{1.5cm} + q_{\alpha}^{n_{\alpha}k_{\alpha}}\tilde{q}_{\alpha}^{k_{\alpha}l_{\alpha}}\dket{\dot{\Dcal}_{\alpha}^{n_{\alpha}}(t)}\dbra{\Ecal_{\alpha}^{l_{\alpha}}(t)} \right)
\text{ . }
\label{eqE}
\end{eqnarray}
Let us then apply the normalization condition for the parameters $q_{\alpha}^{n_{\alpha}m_{\alpha}}$ and $\tilde{q}_{\alpha}^{n_{\alpha}m_{\alpha}}$ in Eq.~(\ref{eqE}). 
By using Eq.~\eqref{ApEqqDelta} in the second term of Eq.~(\ref{eqE}), we then get
\begin{align}
\Lmath_{\text{Gtqd}}(t,t_{0}) & = \sum_{\alpha = 0}^{N-1} \sum _{n_{\alpha} = 1}^{N_{\alpha}}  \sum _{k_{\alpha} = 1}^{N_{\alpha}} \sum _{l_{\alpha} = 1}^{N_{\alpha}} 
\dot{q}_{\alpha}^{n_{\alpha}k_{\alpha}}\tilde{q}_{\alpha}^{k_{\alpha}l_{\alpha}}\dket{\Dcal_{\alpha}^{n_{\alpha}}(t)}\dbra{\Ecal_{\alpha}^{l_{\alpha}}(t)} \nonumber \\
&+\sum_{\alpha = 0}^{N-1} \sum _{n_{\alpha} = 1}^{N_{\alpha}} \dket{\dot{\Dcal}_{\alpha}^{n_{\alpha}}(t)}\dbra{\Ecal_{\alpha}^{n_{\alpha}}(t)}
\text{ . }
\label{eqEE}
\end{align}
Eq.~(\ref{eqEE}) then reproduces the generalized TQD Lindbladian for multi-dimensional Jordan blocks provided by Eq.~\eqref{LG_tqd}.  
Naturally, further conditions can be imposed over  
$q_{\alpha}^{n_{\alpha}m_{\alpha}}$ and $\tilde{q}_{\alpha}^{n_{\alpha}m_{\alpha}}$ for specific transitionless evolutions, such as in the dynamics induced by the standard TQD Lindbladian.

\section{Recovering the standard TQD Lindblad superoperator for a general Jordan decomposition} \label{ApRecTLindGenBlock}

Let us consider $q_{\alpha}^{n_{\alpha}m_{\alpha}}(t) = e^{\int_{t_{0}}^{t} \lambda_{\alpha}(\xi)d\xi}v_{n_{\alpha}m_{\alpha}}(t)$, which means that the adiabatic evolution is mimicked. 
In order to satisfy Eq.~\eqref{ApEqqDelta}, we then have $\tilde{q}_{\alpha}^{n_{\alpha}m_{\alpha}}(t) = e^{-\int_{t_{0}}^{t} \lambda_{\alpha}(\xi)d\xi}\tilde{v}_{n_{\alpha}m_{\alpha}}(t)$. 
Our aim here is to determine the conditions over the parameters $v_{k_{\alpha}l_{\alpha}}$ and $\tilde{v}_{k_{\alpha}l_{\alpha}}$ such that the generalized Lindblad superoperator 
$\Lmath_{\text{Gtqd}}(t)$ reduces to the standard Lindblad superoperator $\Lmath_{\text{Stqd}}(t)$. 
Thus, by computing $\Lmath_{\text{Gtqd}}(t,t_{0})$ according with Eq.~(\ref{eqEE}), we get
\begin{align}
\Lmath_{\text{Gtqd}}(t,t_{0})  
&=\sum_{\alpha = 0}^{N-1} \sum _{n_{\alpha} = 1}^{N_{\alpha}} \dket{\dot{\Dcal}_{\alpha}^{n_{\alpha}}(t)}\dbra{\Ecal_{\alpha}^{n_{\alpha}}(t)} \nonumber \\
&+ \sum_{\alpha = 0}^{N-1} \sum _{n_{\alpha} = 1}^{N_{\alpha}} \sum _{k_{\alpha} = 1}^{N_{\alpha}} \sum _{l_{\alpha} = 1}^{N_{\alpha}} 
\frac{d}{dt}\left[e^{\int_{t_{0}}^{t} \lambda_{\alpha}(\xi)d\xi}v_{n_{\alpha}k_{\alpha}}\right] \times \nonumber\\ 
& \hspace{2.5cm}e^{-\int_{t_{0}}^{t} \lambda_{\alpha}(\xi)d\xi}\tilde{v}_{k_{\alpha}l_{\alpha}}\dket{\Dcal_{\alpha}^{n_{\alpha}}(t)}\dbra{\Ecal_{\alpha}^{l_{\alpha}}(t)} ,
\end{align}
which yields
\begin{align}
\Lmath_{\text{Gtqd}}(t,t_{0})  
 = \sum_{\alpha = 0}^{N-1} \sum _{n_{\alpha} = 1}^{N_{\alpha}} \sum _{k_{\alpha} = 1}^{N_{\alpha}} \sum _{l_{\alpha} = 1}^{N_{\alpha}} &
\lambda_{\alpha}(t)v_{n_{\alpha}k_{\alpha}} \tilde{v}_{k_{\alpha}l_{\alpha}}\dket{\Dcal_{\alpha}^{n_{\alpha}}(t)}\dbra{\Ecal_{\alpha}^{l_{\alpha}}(t)} \nonumber \\
+\sum_{\alpha = 0}^{N-1} \sum _{n_{\alpha} = 1}^{N_{\alpha}} \sum _{k_{\alpha} = 1}^{N_{\alpha}} \sum _{l_{\alpha} = 1}^{N_{\alpha}} &
\dot{v}_{n_{\alpha}k_{\alpha}} \tilde{v}_{k_{\alpha}l_{\alpha}}\dket{\Dcal_{\alpha}^{n_{\alpha}}(t)}\dbra{\Ecal_{\alpha}^{l_{\alpha}}(t)}  \nonumber \\+\sum_{\alpha = 0}^{N-1} \sum _{n_{\alpha} = 1}^{N_{\alpha}} \dket{\dot{\Dcal}_{\alpha}^{n_{\alpha}}(t)&}\dbra{\Ecal_{\alpha}^{n_{\alpha}}(t)} \text{ . }
\label{eqF0}
\end{align}
Now we use the normalization condition between $v_{k_{\alpha}l_{\alpha}}$ and $\tilde{v}_{k_{\alpha}l_{\alpha}}$, provided by $\Vcal_{\text{ad}}(t,t_{0})\Vcal_{\text{ad}}^{-1}(t,t_{0})=\1$. Then, Eq.~(\ref{eqF0}) can be rewritten as 
\begin{align}
\Lmath_{\text{Gtqd}}(t) 
& = \sum_{\alpha = 0}^{N-1} \sum _{n_{\alpha} = 1}^{N_{\alpha}}
\lambda_{\alpha}(t)
\dket{\Dcal_{\alpha}^{n_{\alpha}}(t)}\dbra{\Ecal_{\alpha}^{n_{\alpha}}(t)}  \nonumber \\&+\sum_{\alpha = 0}^{N-1} \sum _{n_{\alpha} = 1}^{N_{\alpha}} \dket{\dot{\Dcal}_{\alpha}^{n_{\alpha}}(t)}\dbra{\Ecal_{\alpha}^{n_{\alpha}}(t)} \nonumber \\&+\sum_{\alpha = 0}^{N-1} \sum _{n_{\alpha} = 1}^{N_{\alpha}} \sum _{k_{\alpha} = 1}^{N_{\alpha}} \sum _{l_{\alpha} = 1}^{N_{\alpha}} 
\dot{v}_{n_{\alpha}k_{\alpha}} \tilde{v}_{k_{\alpha}l_{\alpha}}\dket{\Dcal_{\alpha}^{n_{\alpha}}(t)}\dbra{\Ecal_{\alpha}^{l_{\alpha}}(t)}  \text{ . } \label{H2}
\end{align}
Now, let us work on the last term of Eq.~(\ref{H2}). First, let us consider the original Lindblad superoperator $\Lmath(t)$, as provided by Eq.~(\ref{LindEq}). Its ``quasi"-spectral decomposition reads
\begin{equation}
\Lmath(t) = \sum_{\alpha = 0}^{N-1}\sum _{n_{\alpha} = 1}^{N_{\alpha}} \dket{\Dcal_{\alpha}^{(n_{\alpha}-1)}(t)}\dbra{\Ecal_{\alpha}^{n_{\alpha}}(t)} + \lambda_{\alpha}(t) \dket{\Dcal_{\alpha}^{n_{\alpha}}(t)}\dbra{\Ecal_{\alpha}^{n_{\alpha}}(t)} \text{ . }
\label{eqF}
\end{equation}
We can then rewrite the standard TQD Lindblad superoperator $\Lmath_{\text{Stqd}}(t)= \Lmath(t) + \Lmath_{\text{cd}}(t)$ by using Eqs.~(\ref{eqF}) and (\ref{firstR}), yielding
\begin{align}
\Lmath_{\text{Stqd}}(t) &= \sum_{\alpha = 0}^{N-1}\sum _{n_{\alpha} = 1}^{N_{\alpha}} \dket{\Dcal_{\alpha}^{(n_{\alpha}-1)}(t)}\dbra{\Ecal_{\alpha}^{n_{\alpha}}(t)} + \lambda_{\alpha}(t) \dket{\Dcal_{\alpha}^{n_{\alpha}}(t)}\dbra{\Ecal_{\alpha}^{n_{\alpha}}(t)}  \nonumber \\&\hspace{-0.9cm}+ \sum_{\mu=0}^{N-1}\sum_{n_{\mu}=1}^{N_{\mu}} \left[\dket{\dot{\Dcal}_{\mu}^{n_{\mu}}(t)} \dbra{\Ecal_{\mu}^{n_{\mu}}(t)} - \sum_{k_{\mu}=1}^{N_{\mu}}\Gcal_{\mu\mu}^{n_{\mu}k_{\mu}}(t) \dket{\Dcal_{\mu}^{n_{\mu}}(t)}\dbra{\Ecal_{\mu}^{k_{\mu}}(t)} \right] \text{ , }
\label{H3}
\end{align}
where $\Gcal_{\mu\mu}^{n_{\mu}k_{\mu}}(t) = \dinterpro{\Ecal_{\mu}^{n_{\mu}}(t)}{\dot{\Dcal}_{\mu}^{k_{\mu}}(t)}$. Thus, we can recover the standard Lindblad superoperator $\Lmath_{\text{Stqd}}(t)$ from the generalized Lindblad superoperator 
$\Lmath_{\text{Gtqd}}(t)$ by imposing $\Lmath_{\text{Gtqd}}(t) = \Lmath_{\text{Stqd}}(t)$. From Eqs.~(\ref{H2}) and (\ref{H3}), this is achieved by the requiring
\begin{align}
\sum _{k_{\alpha} = 1}^{N_{\alpha}} \dot{v}_{n_{\alpha}k_{\alpha}}(t) \tilde{v}_{k_{\alpha}l_{\alpha}}(t) &= \delta_{n_{\alpha}(l_{\alpha}-1)} - \dinterpro{\Ecal_{\alpha}^{n_{\alpha}}(t)}{\dot{\Dcal}_{\alpha}^{l_{\alpha}}(t)} \text{ . }
\label{H4}
\end{align}
Hence, Eq.~(\ref{H4}) connects the generalized TQD formalism with the standard counter-diabatic approach.


%

\end{document}